\documentclass[manuscript,screen,acmsmall,authorversion]{acmart}

\usepackage[T1]{fontenc}

\usepackage{graphicx}

\usepackage{afterpage}

\usepackage{amsmath}

\usepackage{url}

\usepackage{makecell}

\usepackage{fontawesome}

\usepackage{enumitem}

\usepackage{multirow}

\usepackage{pdfpages}

\usepackage{diagbox}

\usepackage{tikz}
\newcommand*\emptycirc{\tikz\draw (0,0) circle (1.0ex);} 

\newcommand*\fullcirc{\tikz\fill (0,0) circle (1.0ex);} 

\newcolumntype{P}[1]{>{\centering\arraybackslash}p{#1}}

\graphicspath{{./figures/}}

\AtBeginDocument{%
  \providecommand\BibTeX{{%
    \normalfont B\kern-0.5em{\scshape i\kern-0.25em b}\kern-0.8em\TeX}}}

\setcopyright{none}
\acmJournal{JDIQ}
\acmYear{2023} 
\acmVolume{1} 
\acmNumber{1} 
\acmPrice{15.00}
\acmDOI{10.1145/3623640}

\begin{document}

\title[Validating Synthetic Usage Data in Living Labs]{Validating Synthetic Usage Data in Living Lab Environments}



\author{Timo Breuer}
\orcid{0000-0002-1765-2449}
\affiliation{%
\institution{TH K{\"o}ln - University of Applied Sciences}
\country{Germany}}
\email{timo.breuer@th-koeln.de}


\author{Norbert Fuhr}
\orcid{0000-0002-0441-6949}
\affiliation{%
\institution{University of Duisburg-Essen}
\country{Germany}}
\email{norbert.fuhr@uni-due.de}


\author{Philipp Schaer}
\orcid{0000-0002-8817-4632}
\affiliation{%
\institution{TH K{\"o}ln - University of Applied Sciences}
\country{Germany}}
\email{philipp.schaer@th-koeln.de}

\renewcommand{\shortauthors}{Timo Breuer, Norbert Fuhr, Philipp Schaer}

\begin{abstract}
Evaluating retrieval performance without editorial relevance judgments is challenging, but instead, user interactions can be used as relevance signals. Living labs offer a way for small-scale platforms to validate information retrieval systems with real users. If enough user interaction data are available, click models can be parameterized from historical sessions to evaluate systems before exposing users to experimental rankings. However, interaction data are sparse in living labs, and little is studied about how click models can be validated for reliable user simulations when click data are available in moderate amounts.
    
This work introduces an evaluation approach for validating synthetic usage data generated by click models in data-sparse human-in-the-loop environments like living labs. We ground our methodology on the click model's estimates about a system ranking compared to a reference ranking for which the relative performance is known. Our experiments compare different click models and their reliability and robustness as more session log data becomes available. In our setup, simple click models can reliably determine the relative system performance with already 20 logged sessions for 50 queries. In contrast, more complex click models require more session data for reliable estimates, but they are a better choice in simulated interleaving experiments when enough session data are available. While it is easier for click models to distinguish between more diverse systems, it is harder to reproduce the system ranking based on the same retrieval algorithm with different interpolation weights. Our setup is entirely open, and we share the code to reproduce the experiments.
\end{abstract}




\maketitle

\section{Introduction}

One of the primary goals in IR evaluation is to find the best-performing system, i.e., to identify the relative ordering of retrieval systems by the effectiveness referred to as the \textit{system ranking} in the following. For many decades, these IR benchmarks have been conducted according to the Cranfield paradigm \cite{DBLP:conf/sigir/Cleverdon91}, for which the data curation of the underlying test collections comes at a high cost and is usually only feasible as part of larger community efforts like shared tasks at CLEF, FIRE, NTCIR, or TREC  \cite{DBLP:journals/ftir/Sanderson10}.

A completely different approach to curating relevance feedback data for IR systems is made possible by online experiments \cite{DBLP:journals/ftir/HofmannLR16,DBLP:conf/kdd/KohaviDFWXP13}. In this case, user interaction feedback is used to estimate the relevance of the search results. Large-scale web search companies can rely on an abundance of such data but cannot share it due to privacy concerns and business interests \cite{DBLP:conf/cikm/CraswellCMYB20}. For this reason, there are few datasets covering document collection, user interaction data, and the corresponding search engine result pages (SERPs).

\begin{figure}[!t]
    \includegraphics[width=\textwidth]{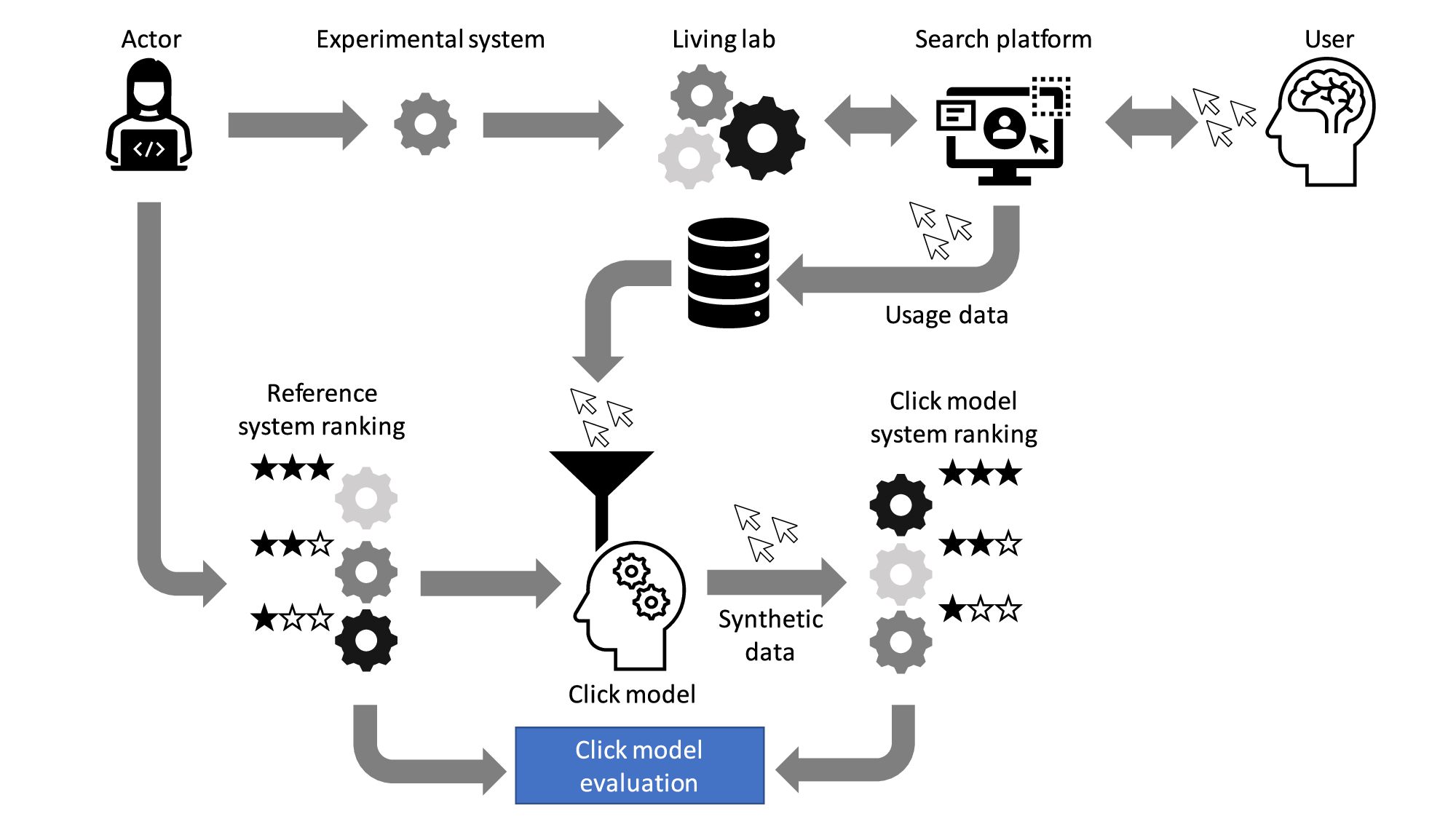}
    \caption[]{Click model evaluations based on system rankings and logged user interaction data from living labs.}
    \label{fig:click.model.evaluation}
\end{figure}

In order to pave the way for experimental evaluations in small- to mid-scale user data environments, the living lab paradigm \cite{DBLP:journals/jdiq/JagermanBR18} was introduced. As part of these efforts, small and domain-specific search services with different applications like product or academic search opened their infrastructures for researchers who are able to evaluate their IR systems in online experiments with user interaction data. Opposed to A/B experiments, which only deliver meaningful results with large amounts of user data, previous living labs \cite{DBLP:conf/clef/SchuthBK15,DBLP:conf/cikm/GingstadJB20,DBLP:conf/clef/SchaerBCWST21a} implemented the experimental design based on interleavings. The general idea is to combine ranking lists of two or more retrieval systems, show the interleaved ranking to users, and let them decide on the better-performing system by their click decision based on their relative preference.

Earlier works concluded that user interaction data in living labs is sparse \cite{DBLP:conf/clef/SchuthBK15,DBLP:journals/jdiq/JagermanBR18,DBLP:conf/clef/SchaerBCWST21a}. While there is a need to validate laboratory or system-oriented experiments in the real world, the corresponding experiments come with the risk of harming the user experience. The risk is even higher for small and domain-specific search services, and it is a desideratum to keep the online time of experimental systems short while having insights about their usefulness. As a way out, synthetic usage data can be considered a possibility to account for a user-oriented evaluation without the risk of exposing real users to bad search results.

User interactions like clicks are alternative \textit{relevance signals} or proxies that could be used for estimating the system performance from a different perspective \cite{DBLP:conf/sigir/ChenZL0M17}. If enough user interaction data are available, it is possible to parameterize click models that can be used for generating synthetic user interactions. These click models bear the potential to replace real users in living labs when evaluating highly experimental systems. As user interaction data are low, it is of high interest to make an estimate of how much data are required for a robust parameterization of the click model to use it in a reliable way when generating synthetic interactions. 

This work is about validating synthetic usage data of click models in data-sparse environments like living labs. Figure \ref{fig:click.model.evaluation} illustrates the evaluation task, where the \textit{actor} is interested in validating the retrieval effectiveness and usefulness of an \textit{experimental system} in the real world. \textit{Living labs} offer a gateway to user experiments. Once submitted to the living lab infrastructure, the experimental system can be deployed on the backend of \textit{search platforms}, which, in turn, can provide users with results in interleaving experiments. User interaction feedback data like clicks are logged and sent to a central database of the living lab infrastructure.

The actor usually has the choice of different systems or configurations, and not all of them are worth being validated in a user experiment. As a solution, the actor could select suitable systems for online experiments in a pre-assessment step based on user simulations. In this case, a click simulator is used to separate good-performing systems from the rest. As the user interactions are continuously logged, it is possible to update the click model's parameters with new log data. But how does the actor know when the click model is parameterized well enough such that it can be used for good user simulations?

We propose an evaluation approach in which the click model has to decide about the relative system performance, i.e., the system ranking. According to this method, the actor provides the click models with a \textit{reference system ranking}, for which the relative system performance is known in advance with high confidence. Based on its click decisions and the generated click data, the model itself also produces a \textit{click model system ranking}, which can be compared to the reference system ranking. If the click model returns the correct system ranking, it can be considered a suitable user simulator that generates meaningful synthetic usage data.

Recently, several datasets, for which the ground truth relevance of query document pairs was inferred from click signals, were released \cite{DBLP:conf/sigir/RekabsazLSBE21,DBLP:conf/sigir/ZhengFLLZM18,DBLP:conf/cikm/ZhangLMT18}. More specifically, multi-graded relevance labels were derived from the click-through rate and threshold values \cite{DBLP:conf/sigir/RekabsazLSBE21}. However, we note that clicks are not a direct substitute for editorial relevance judgments like they are made for IR test collections, as found in a previous study \cite{DBLP:conf/wsdm/KampsKT09}. In addition, finding reasonable threshold values to make multi-graded labels from the click-through rate is not well studied and can be critical as the threshold criteria might differ across queries/topics \cite{DBLP:conf/ecir/ScholerSBT08}.

Instead, we see user clicks as alternative \textit{relevance signals} that require a different evaluation approach. For this reason, the work at hand investigates the problem of evaluating the relative system performance based on click models in the absence of editorial relevance judgments. More specifically, we evaluate to which extent click models can determine the relative system performance by the Log-Likelihood of the click probability and, afterward, in simulated interleaving experiments. 

We analyze the reliability and robustness of click models for estimating the relative system performance by evaluating the click models' system ranking over an increasing amount of queries and click logs. More specifically, we compare the Document-based Click-Through Rate Model (DCTR) to the Dependent Click Model (DCM) and the Simplified Dynamic Bayesian Network Model (SDBN), which embed the continuation probability and the notion of satisfying clicks. 

Furthermore, we include two types of system rankings. The first is based on different lexical retrieval methods (LRM), whereas the second is made from interpolated retrieval methods (IRM). For both the Log-Likelihood and the interleaving experiments, we determine the correlation with a reference ranking by Kendall's $\tau$. More precisely, we give answers to the following research questions:

\begin{enumerate}[label=\textbf{RQ\arabic*},leftmargin=10mm]
    \item \emph{Can click models reproduce system rankings?} 
    \item \emph{Do continuation and satisfaction probabilities in click models improve the simulation quality?}
    \item \emph{How does the type of system ranking impact the outcomes of simulated interleaving experiments?} 
\end{enumerate}

\textbf{RQ1} addresses the general plausibility of the introduced evaluation approach. The focus of \textbf{RQ2} is the comparison of DCTR to more complex models. In particular, DCM and SDBN have a less abstract user model than DCTR. Besides the attractiveness of search results, they account for the click sequence and whether there are satisfied clicks. \textbf{RQ3} addresses two different types of system rankings that could be compared. While the LRM ranking is composed of more distinct systems, the IRM ranking is based on the same system with different interpolation weights. Besides the answers to our research questions, the contributions of this work are as follows:

\begin{itemize}
    \item We \textbf{introduce an evaluation approach} for validating synthetic usage data generated by click models in data-sparse \textit{human-in-the-loop} environments like living labs,
    \item \textbf{compare two different system rankings}, including lexical-based systems and the same system with different interpolation weights to evaluate the proposed methodology, 
    \item \textbf{compare three different click models}, including DCTR, DCM, and SDBN,
    \item \textbf{validate the proposed methodology} by simulated interleaving experiments with state-of-the-art Transformer-based rankings,
    \item \textbf{provide an open and fully reproducible experimental setup} including open-source code and open data.\footnote{\faGithub \ \url{https://www.github.com/irgroup/validating-synthetic-usage-data}}
\end{itemize}

The remainder is structured as follows. Section \ref{sec:related_work} reviews the related work about living labs, user simulations, and click models. Section \ref{sec:methodology} outlines the methodology and the experimental setup, whereas Section \ref{sec:experimental_evaluations} presents the experimental evaluations. Section \ref{sec:answers} gives answers to our research questions. Finally, Section \ref{sec:conclusions} discusses the results and concludes. 

\section{Related Work and Background}
\label{sec:related_work}

This section reviews the related work about living lab experiments, briefly summarizes relevant work about user simulations, and finally, provides the fundamentals of click models.

\subsection{Living Labs}

The principle of the living lab paradigm within the scope of shared tasks can be described as follows. Participants contribute their experimental systems or sometimes only the pre-computed outputs to the living lab platform, which connects participants and their experimental systems on the one side with the connected search services on the other side. Users can then be provided with the experimental results upon request, and their interactions will be logged in order to evaluate or improve the experimental systems.

One of the earlier works that mentioned the idea of a ``living laboratory'' was made by Kelly et al.~\cite{DBLP:journals/computer/KellyDP09} and dates back to 2009. The idea was picked up by Azzopardi and Balog~\cite{DBLP:conf/clef/AzzopardiB11}, who made the first proposal for a living lab architecture in 2011. In 2013, a workshop dedicated to living labs discussed several requirements and extensions of the living lab paradigm \cite{DBLP:conf/cikm/BalogEKKS13} followed by the first implementation of the living lab architecture for ad-hoc IR experiments in 2014 \cite{DBLP:conf/cikm/BalogKS14}. Finally, the first living lab for ad-hoc retrieval was held at CLEF in 2015 and was continued in a second iteration in 2016 \cite{DBLP:conf/clef/SchuthBK15}. The same organizers were also involved in the Open Search track at TREC in 2016 and 2017 \cite{DBLP:journals/jdiq/JagermanBR18}. NEWSREEL was the first living lab for real-time news recommendations and ran from 2014 until 2017 \cite{DBLP:conf/clef/HopfgartnerKLPBH14,DBLP:conf/iiix/BrodtH14}. 
More recent living lab implementations are not specifically tailored for shared tasks but have a domain-specific focus. Some recent examples include APONE~\cite{DBLP:conf/desires/Marrero18,DBLP:conf/sigir/MarreroH18} and arXivDigest~\cite{DBLP:conf/cikm/GingstadJB20}. APONE is a living lab platform designed for A/B tests focusing on evaluating user interfaces. As it builds upon the PlanOut language \cite{DBLP:conf/www/BakshyEB14}, it allows designing the experiments by scripting them. arXivDigest is a recommendation service for research articles based on personalized email updates on recent publications from arXiv's computer science repositories. After registration, an interest profile helps to find adequate recommendations, and feedback is provided with the help of clicked URLs in the personalized mail. Besides arXivDigest, Beel et al.~\cite{DBLP:conf/ecir/BeelCKDK19} also provide a living lab platform for scholarly recommendations.

More recently, Schaer et al.~\cite{DBLP:conf/isiwi/BreuerS21,DBLP:conf/clef/SchaerBCWST21a} presented a novel infrastructure design for living labs. The infrastructure was tailored explicitly for shared task collaborations and was the backbone of the LiLAS lab at CLEF in 2021. One of the substantial improvements over earlier living lab attempts is the possibility of submitting the entire experimental system instead of submitting pre-computed results only, which addresses the shortcoming of pre-computed results in earlier living labs \cite{DBLP:journals/jdiq/JagermanBR18}. 

By using only pre-computed results for selected queries, the experiments are artificially shrunk to a subset of queries. Even more, they may be outdated quickly, which can become critical in e-commerce settings. Instead, Schaer et al.~\cite{DBLP:conf/isiwi/BreuerS21,DBLP:conf/clef/SchaerBCWST21a} envision a dockerized retrieval system that can dynamically be updated and deliver results for arbitrary queries.

Participants of the shared task provide systems with retrieval and recommendation algorithms in the form of micro-services that can be deployed on purpose in a reproducible way. The infrastructure builds upon Docker and its containerization technology to make this possible. An additional central component is Git and the integration of the web service GitHub, facilitating the experimental components' software versioning, transparency, and reproducibility. 

Once the systems are implemented, the experimenters prepare them with Docker containers. More specifically, they prepare a dockerizable source code repository, and after registration, each dockerized system can be integrated into a single multi-container application. Multiple systems from possibly different experimenters are combined, which means that the administrators at the search platform do not have to set up individual systems but rather can rely on complete replicas of all submitted systems once the multi-container application is running.

Each search platform deploys an instance of the multi-container application on its backend servers. Queries from users will then be redirected from the search interfaces to the individual experimental systems. Upon request, experimental search results are returned, and the search platform is supposed to log user interaction data, which eventually is sent to a central server of the infrastructure, where it is stored and can be used for further analysis, training, and optimization of the experimental systems.

While Schaer et al. conclude that their infrastructure design overcomes the bottleneck of pre-computed queries, there were still moderate amounts of logged user interaction data that only partially allowed for statistical significance tests \cite{DBLP:conf/clef/SchaerBCWST21a}. Small- and mid-scale search platforms generally have moderate user traffic, and relevance feedback is generally sparse \cite{DBLP:journals/jdiq/JagermanBR18}. 

However, it is common practice to reuse historical session logs to evaluate new ranking methods before exposing them to real users \cite{DBLP:conf/wsdm/LiKZ15}, either to avoid harming the user experience or to reduce online time in order to increase the rate at which new experiments can be conducted \cite{DBLP:conf/sigir/DrutsaGKKSY19,DBLP:conf/sigir/KharitonovMSO15}. 

As an alternative to A/B experiments, which only deliver meaningful results with a large amount of user data, experimental systems can be deployed in interleaving experiments like it is often done in living lab environments \cite{DBLP:journals/jdiq/JagermanBR18,DBLP:conf/clef/SchuthBK15,DBLP:conf/cikm/GingstadJB20,DBLP:conf/clef/SchaerBCWST21a}. The general idea is to combine ranking lists of two or more retrieval systems and let users decide on the better-performing system by their click decisions based on the relative preference. There exist different interleaving strategies like probabilistic interleaving \cite{DBLP:journals/tois/HofmannWR13}, multileaving \cite{DBLP:conf/wsdm/SchuthOWR16}, preference-based balanced \cite{DBLP:conf/cikm/HeZL09}, or temporal interleaving \cite{DBLP:conf/sigir/QianLR16}, but the team draft interleaving \cite{DBLP:conf/cikm/RadlinskiKJ08} is more commonly used and also studied in this work.

While interleaving reduces the risk of returning poor search results by combining experimental rankings with a reasonable baseline ranking, there is still the risk of harming the user experience. Preferably, promising systems should be identified before deploying them in online experiments. A viable solution for pre-assessments is user simulation, which will be described next.

\subsection{User Simulations}

The most prominent user model in system-oriented evaluations implies that the user formulates a single query for a given information need, scans the entire result list up to a fixed rank, and judges the relevance of each item independent of any context knowledge, e.g., from previously seen results \cite{DBLP:phd/ethos/Maxwell19,DBLP:journals/sigir/BalogMTZ21}. However, depending on the IR measure, additional assumptions about the underlying user model are made as part of the evaluations. For instance, nDCG \cite{DBLP:journals/tois/JarvelinK02} discounts later items in the ranking by log-harmonic weights and, thus, simulates the user's persistence. Similarly, the RBP also allows defining the user's persistence \cite{DBLP:journals/tois/MoffatZ08}. 

Carterette~\cite{DBLP:conf/sigir/Carterette11} introduced a coherent framework for model-based measures. Similarly, Moffat et al.~\cite{DBLP:journals/tois/MoffatBST17} introduced the C/W/L framework to describe a family of parameterizable evaluation measures that account for the user browsing behavior by formalizing the conditional continuation probability of examining items in the ranking list. Both of these frameworks are able to describe conventional measures like nDCG, AP, or RBP but also allow for the analysis of derived variants. While all of these measures allow for a principled system-oriented evaluation over different topics with certain assumptions about the user behavior, they are still a strong abstraction of how the user interacts with the search system, and the user behavior has a somewhat static notion. 

Based on the idea of extending the underlying user model of system-oriented experiments, simulations make it feasible to evaluate retrieval systems with regards to more \textit{dynamic} user interactions. For instance, earlier seen retrieval results can be exploited for more diverse query formulations over multiple result pages, situational clicks, relevance decisions, and diverging browsing depths \cite{DBLP:conf/ictir/CarteretteBZ15}. Simulated IR experiments date back to the early 1980s \cite{DBLP:conf/sigir/TagueNW80,DBLP:conf/sigir/TagueN81}, but more recently, several frameworks and user models were introduced \cite{DBLP:conf/cikm/BaskayaKJ13,DBLP:conf/iiix/ThomasMBS14,DBLP:conf/ictir/CarteretteBZ15,DBLP:conf/cikm/MaxwellA16,DBLP:conf/sigir/MaxwellA16,DBLP:journals/ir/PaakkonenKKAMJ17,DBLP:conf/ictir/ZhangLZ17}. Inspired by the user models of Baskaya et al.~\cite{DBLP:conf/cikm/BaskayaKJ13} and Thomas et al.~\cite{DBLP:conf/iiix/ThomasMBS14}, Maxwell and Azzopardi~\cite{DBLP:conf/cikm/MaxwellA16,DBLP:conf/sigir/MaxwellA16} introduced the \textit{Complex Searcher Model}. Carterette et al.~\cite{DBLP:conf/ictir/CarteretteBZ15} proposed the idea of \textit{Dynamic Test Collections}, and Pääkönen et al.~\cite{DBLP:journals/ir/PaakkonenKKAMJ17} introduced the \textit{Common Interaction Model}. Zhang et al.~\cite{DBLP:conf/ictir/ZhangLZ17} recently introduced another search simulation framework. As a special type of user simulation, the focus of click models is generating click interactions with the ranking list. 

In order to compare the fidelity of user simulation, Labhishetty and Zhai~\cite{DBLP:conf/sigir/LabhishettyZ21,DBLP:conf/ecir/LabhishettyZ22} introduced the Tester-based approach. The key idea is based on the definition of \textit{Testers} that are composed of single retrieval systems for which the relative retrieval effectiveness is known. The user simulator is evaluated by how well it can identify the correct relative retrieval effectiveness.

\subsection{Click models}

In contrast to explicit editorial relevance judgments of test collections, click signals, or user interactions in general, are a more implicit form of relevance feedback \cite{DBLP:conf/sigir/JoachimsGPHG05}, which is often used to improve the quality of search results \cite{DBLP:conf/sigir/AgichteinBD06}. Generally, it is controversially discussed how user interactions like clicks can reflect topical relevance. While several studies suggest that improved system performance does not directly translate into better user performance \cite{DBLP:conf/sigir/HershTPCKSO00,DBLP:conf/sigir/TurpinH01,DBLP:conf/sigir/TurpinH02,DBLP:conf/adc/TurpinH04,DBLP:conf/sigir/TurpinS06,DBLP:conf/wsdm/KampsKT09}, some works concluded that user and system metrics correlate under certain constraints \cite{DBLP:conf/sigir/Zobel98,DBLP:conf/sigir/JoachimsGPHG05,DBLP:conf/sigir/SandersonPCK10,DBLP:conf/chiir/JiangA16,DBLP:conf/sigir/ChenZL0M17}. It is beyond the scope of this work to draw any conclusions about how clicks correlate with topical relevance judgments and we consider clicks as an alternative that can be used as a proxy when it is not feasible to have editorial relevance judgments. 

While earlier click models mostly differ by the predefined rules that make assumptions about the underlying user behavior \cite{DBLP:series/synthesis/2015Chuklin,DBLP:conf/clef/GrotovCMSXR15}, several improved models were introduced, accounting for clicks on multiple result pages, and aggregated search \cite{DBLP:conf/cikm/ChuklinSHSR13,DBLP:journals/tois/ChuklinSZR15}, embedding time awareness by accounting for dwell times and timestamps between click sequences \cite{DBLP:journals/tois/LiuXWNZM17}, or omitting predefined rules by replacing them with neural vector states learned from user logs \cite{DBLP:conf/www/BorisovMRS16}, or embedding global and local click models into a framework for better personalization \cite{DBLP:conf/www/Zhang0MXZMT22}. Click models can be distinguished by the parameter estimation, which is either done by maximum likelihood estimation (MLE) or the expectation-maximization (EM) algorithm, which has been improved for more efficiency \cite{DBLP:conf/www/KhandelMYV22} and online retraining \cite{DBLP:conf/cikm/MarkovBR17}. Suppose both clicks and editorial relevance judgments are available. In that case, it is possible to turn click models into information retrieval metrics \cite{DBLP:conf/sigir/ChuklinSR13} or to make new relevance labels for previously unjudged documents \cite{DBLP:conf/nips/CarteretteJ07,DBLP:conf/www/OzertemJD11}.

The quality of click models is often evaluated by the Log-Likelihood and Perplexity \cite{DBLP:conf/ictir/MalkevichMMR17}, but also other reliability measures exist \cite{DBLP:conf/ictir/MaoCLZM19}. In previous work, click models have mainly been evaluated on semi-public web search datasets, e.g., from Yahoo! \cite{DBLP:conf/sigir/CarteretteC18,DBLP:conf/kdd/LiAKMVW18,DBLP:conf/www/OzertemJD11,DBLP:conf/cikm/PiwowarskiZ07} or Yandex \cite{DBLP:conf/cikm/ChuklinSHSR13,DBLP:conf/clef/GrotovCMSXR15}, in which the SERPs are anonymized and the underlying web corpus is fully or partially private. To the best of our knowledge, we are the first to evaluate simulated interleaving experiments with a completely open and transparent experimental setup. 

\section{Methodology and Evaluation Setup}
\label{sec:methodology}

As described in the introduction of this work, our overall methodology aims to validate the simulation quality of click models in interaction data-sparse environments like living labs. The key idea is to validate the click model by its ability to identify the correct system ranking, for which we know the relative performance of each system with high confidence in advance. To better understand how much user interaction data are required for a reasonable parameterization of the click model, we parameterize and subsequently evaluate the click model over an increasing amount of session logs. The performance estimates of the click model are either based on the Log-Likelihood or on the highest click probability, which is used in simulated interleaving experiments. The click model system ranking is compared to the reference ranking with the help of Kendall's $\tau$, which determines the rank correlation.

In the following, we describe the two types of system rankings and the corresponding single systems that will be used as the reference system rankings to which the performance estimates of the click models are compared (cf. \ref{sec:experimental_systems}). Afterward, we describe and compare the three click models by their attractiveness and examination probabilities (cf. \ref{sec:click_models}). In comparison, the click models mainly differ by how the examination probability is determined, and we discuss the corresponding assumptions about the concepts of satisfaction and continuation by an illustrative toy example. Furthermore, we describe our experimental setup that is based on the TripClick dataset (cf. \ref{sec:dataset}) and introduce the evaluation measures (cf. \ref{sec:evaluation_measures}). Finally, we provide details about the implementation and hardware (cf. \ref{sec:implementation}).

\subsection{Experimental Systems}
\label{sec:experimental_systems}

In our experiments, we include two types of system rankings, and selecting them is motivated by the Tester-based approach by Labhishetty and Zhai \cite{DBLP:conf/sigir/LabhishettyZ21,DBLP:conf/ecir/LabhishettyZ22}. According to them, a user simulator (in this study, it is the click model) can be validated by its ability to distinguish the retrieval performance of methods for which we know the relative system effectiveness with high confidence or based on heuristics. For instance, by experience, we can safely assume that BM25 is more effective than ranking documents by the term frequency. The first system ranking is based on \textbf{L}exical \textbf{R}etrieval \textbf{M}ethods (\textbf{LRM}) and is defined by

\begin{equation*}
    \mathrm{DFR} \chi^2 \succ \mathrm{BM25} \succ \mathrm{Tf} \succ \mathrm{Dl} \succ \mathrm{Null}.
\end{equation*}

More specifically, it is composed of the following five methods (in decreasing order of hypothesized effectiveness), including (1) the DFR $\chi^2$ model \cite{DBLP:conf/ecir/Amati06}, which is a (free from parameters) DFR method based on Pearson's $\chi^2$ divergence, (2) the BM25~\cite{DBLP:journals/ftir/RobertsonZ09} method, (3) the term frequency (Tf) of the query terms in the document, (4) the query-agnostic method based on document length (Dl), and (5) a method that assigns score values of zero (Null).

In contrast, the second system ranking is composed of an \textbf{I}nterpolated \textbf{R}etrieval \textbf{M}ethod (\textbf{IRM}) with different interpolations between a reasonable and a less effective retrieval method, which gives us more control over the effectiveness by weighting the influence of the less effective retrieval method. In our experiments, we combine the DFR ranking method with the ranking criterion based on document length (Dl) and determine the ranking score for a document-query pair ($d,q$) as follows: 

\begin{equation}
    \mathrm{score}(d,q) = \rho \cdot \mathrm{score\textsubscript{Dl}}(d,q) + (1 - \rho) \cdot  \mathrm{score\textsubscript{DFR}}(d,q).
\end{equation}

\begin{figure}[!t]
    \includegraphics[width=\textwidth]{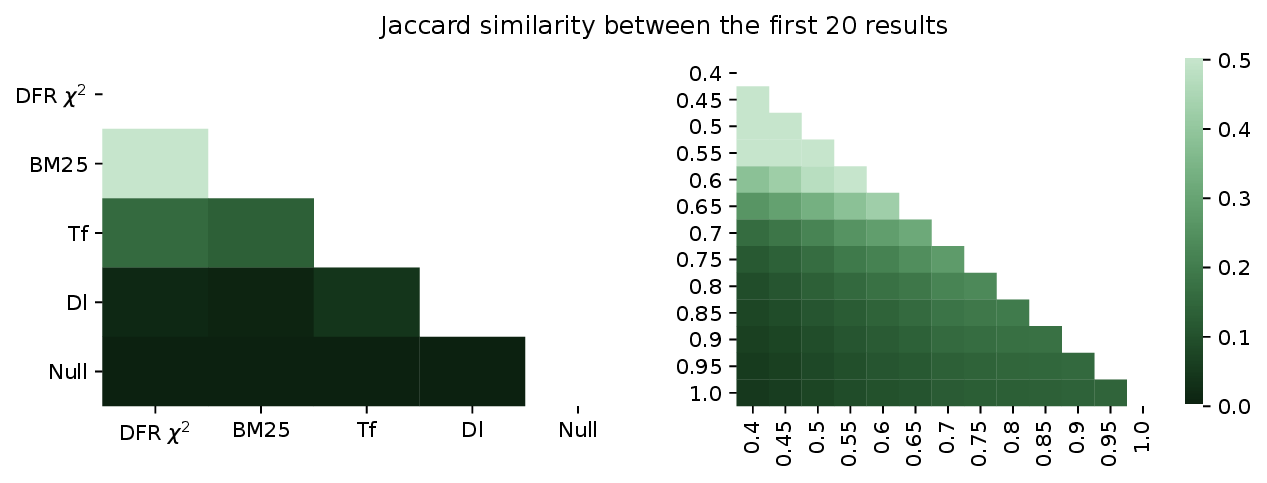}
    \caption[]{Jaccard similarity between the first 20 documents of the 50 head queries for the \textbf{LRM} (left) and \textbf{IRM} (right) system rankings. The Jaccard index is determined based on the document identifiers of both rankings.}
    \label{fig:jacc}
\end{figure}

By increasing $\rho$, we deteriorate the ranking results in a systematic but also more subtle way, which better simulates incremental and less invasive changes to an existing search platform in an online experiment.\footnote{We exclude interpolations with $\rho < 0.4$ to cover a similar score range of the Jaccard similarity for the LRM and IRM rankings, as shown in Figure \ref{fig:jacc}.} The resulting IRM ranking is defined by 

\begin{equation*}
    \mathrm{IRM}_{\rho_1} \succ \mathrm{IRM}_{\rho_2} \succ \cdots \succ \mathrm{IRM}_{\rho_n} 
\end{equation*}

where $\mathrm{IRM}_{\rho}$ denotes a single system in the ranking, and ${\rho_1} < {\rho_2} < ... < {\rho_n}$, i.e., a lower interpolation weight $\rho$ is supposed to result in a more effective retrieval system. We acknowledge that the intuition of the relative IRM system ranking is based on a weak heuristic that is only valid for this particular combination of retrieval methods (as will be evaluated in Subsection~\ref{sec:evaluation_measures}). Generally, it cannot be guaranteed that a higher interpolation parameter, giving more weight to the inferior ranking method, will decrease effectiveness. Especially if the difference in effectiveness is moderate, the linear combination of reasonable retrieval methods can improve the results, as exploited in many data fusion experiments. However, in our settings, we ensure a decrease in effectiveness by implementing the inferior retrieval method with a query-agnostic ranking criterion that will likely deteriorate the ranking.

When comparing the LRM and IRM rankings, the LRM ranking has more diverse document rankings, as shown in Figure \ref{fig:jacc}. The heatmaps compare the first 20 results of the document rankings for the 50 most frequent queries of the dataset described in Section \ref{sec:dataset} between the combinations of the different systems by the Jaccard similarity.
Given the rankings of the two systems, we compare the corresponding document sets by the Jaccard similarity. The higher the Jaccard index, the more similar the two document sets. We note that a perfect Jaccard index of 1.0 could be achieved with the same document sets but different rankings, i.e., the documents in both rankings do not need to be in the same order. This evaluation focuses \textit{diversity} in the document rankings. For the evaluation of rank correlations, Kendall's $\tau$ or the Rank-biased Overlap (RBO)~\cite{DBLP:journals/tois/WebberMZ10} should be preferred (cf. Subsection~\ref{sec:evaluation_measures}).
Except for the comparison of DFR and BM25, most of the LRM combinations are quite dissimilar. In comparison, the IRM systems with different interpolation weights cover a similar score range but have a more gradual transition of the Jaccard similarity over the different combinations of weight pairs. The LRM ranking includes fewer but has more distinct systems. In contrast, the IRM ranking is based on more similar document rankings but also more systems, which means that changing the rank position of a single system would result in less severe changes in Kendall's $\tau$ as compared to changes in the LRM ranking. 

\subsection{Click Models}
\label{sec:click_models}

In the following, we review the analyzed click models~\cite{DBLP:series/synthesis/2015Chuklin}, which are based on probabilistic modeling of the underlying user behavior as opposed to other models based on neural networks \cite{DBLP:conf/www/BorisovMRS16,Zhang2023}. All of them can only estimate the click probability of query-document pairs that were available during the parameter optimization. Given a document ranking, a click model estimates the probability $P\left(C_{d}=1 \mid \mathbf{C}_{<r}\right)$ of a click $C_{d}$ on the document $d$ considering earlier clicks $\mathbf{C}_{<r}$ before the rank $r$ by

\begin{equation}
    P\left(C_{d}=1 \mid \mathbf{C}_{<r}\right)=P\left(C_{d}=1 \mid E_{r}=1\right) \cdot P\left(E_{r}=1\right)=\alpha_{d q} \varepsilon_{r}
\end{equation}

where the probability $P\left(C_{d}=1 \mid E_{r}=1\right)$ depends on the probability $P\left(E_{r}\right)$ that the document is examined. Thus, the click probability of a document $d$ can be decomposed into the \textit{attractiveness} $\alpha_{d q}$ of the query-document pair $(d,q)$ and the \textit{examination} probability $\varepsilon_{r}$. The attractiveness of all click models in this study is given by 

\begin{equation}
    \alpha_{d q}=\frac{1}{\left|\mathcal{S}_{d q}\right|} \sum_{s \in \mathcal{S}_{d q}} c_{d}^{(s)}
\end{equation}

and only differs by the set of sessions $\mathcal{S}_{d q}$. In this work, a session $s$ covers a single query, a corresponding SERP with ranked items, and multiple clicks. Unlike other works, the analyzed click models do not consider multi-query sessions. We acknowledge the simplified understanding of a session that contrasts other user-oriented studies that, for instance, consider query reformulations for the same information need. The DCTR model determines the click probability solely by the ratio of clicks on a document $d$ and how often it has been shown to users for a query $q$. The attractiveness is determined over all available sessions where $q$ and $d$ occur. The examination probability of DCTR for the document at the next rank ($r+1$) is defined as  

\begin{equation}
    \varepsilon_{r+1}=1
\end{equation}

i.e., the click model does not consider the context of other documents and the notion of \textit{satisfaction}. In comparison, both click models DCM and SDBN extend the \textit{cascade model} \cite{DBLP:conf/wsdm/CraswellZTR08} and determine the attractiveness by considering sessions with documents before the last-clicked document at rank $l$ in a particular session, assuming that the user continued the search after having clicked unsatisfying results and documents beyond $l$ were not observed by the user. The set of sessions is defined as

\begin{equation}
    \mathcal{S}_{d q}=\left\{s_{q}: d \in s_{q}, r \leq l\right\}.
\end{equation}

In order to account for the satisfaction of clicks, the DCM introduces the continuation probability $\lambda_{r}$ determined by the ratio between the total number of sessions with clicks at rank $r$ that were not the last click in a session (denoted as $\mathcal{I}(r \neq l)$) and the total number of sessions in which rank $r$ was logged $\left|\mathcal{S}_{r}\right|$. The continuation probability $\lambda_{r}$ is defined as 

\begin{equation}
    \lambda_{r}=\frac{1}{\left |\mathcal{S}_{r}\right |} \sum_{s \in \mathcal{S}_{r}} \mathcal{I}(r \neq l).
\end{equation}

The examination probability $\varepsilon_{r+1}$ of DCM is then defined as

\begin{equation}
    \varepsilon_{r+1}=c_{r}^{(s)} \lambda_{r}+\left(1-c_{r}^{(s)}\right) \frac{\left(1-\alpha_{d q}\right) \varepsilon_{r}}{1-\alpha_{d q} \varepsilon_{r}}
\end{equation}

where $c_{r}^{(s)}$ denotes the probability of a click being observed at rank $r$ in a session $s$. Similarly, the SDBN model embeds the satisfaction probability by the parameter $\sigma_{d q}$ but instead, it accounts for the total number of sessions with the last clicks (denoted as $\mathcal{I}\left(r_{d}^{(s)}=l\right)$) in reference to the total number of sessions $\mathcal{S}_{d q}^{\prime}$ in which the document $d$ is clicked at a rank before or equal to $l$. The satisfaction probability $\sigma_{d q}$ is defined as 

\begin{equation}
    \sigma_{d q}=\frac{1}{\left|\mathcal{S}_{d q}^{\prime}\right|} \sum_{s \in \mathcal{S}_{d q}^{\prime}} \mathcal{I}\left(r^{(s)}=l\right)
\end{equation}

where the corresponding set of sessions $\mathcal{S}_{d q}^{\prime}$ is defined by

\begin{equation}
    \mathcal{S}_{d q}^{\prime}=\left\{s_{q}: d \in s_{q}, r \leq l, c_{d}^{(s)}=1\right\}.
\end{equation}

The examination probability $\varepsilon_{r+1}$ of SDBN is then defined as

\begin{equation}
    \varepsilon_{r+1}=c_{r}^{(s)} \left(1-\sigma_{d q}\right)+\left(1-c_{r}^{(s)}\right) \frac{\left(1-\alpha_{d q}\right) \varepsilon_{r}}{1-\alpha_{d q} \varepsilon_{r}}.
\end{equation}

For the sake of better comparability, Table \ref{tab:cm} provides an overview of how the click models' attractiveness and examination probabilities are determined. For all three click models, the parameters are derived from observable variables, e.g., via the MLE algorithm.

\begin{table*}[!t]
    \caption[]{Click models: the examination probability $\varepsilon_{r+1}$ of DCM and SDBN depends on $c_{r}^{(s)}$ that denotes the probability of a click being observed at rank $r$ in a session $s$.}
    \label{tab:cm}
    \centering
    \begin{tabular}{|l|l|l|}
    \hline 
    Click model & $\mathcal{S}_{d q}$ of attractiveness $\alpha_{d q}$ & Examination probability $\varepsilon_{r+1}$ \\
    \hline
    DCTR \cite{DBLP:conf/wsdm/CraswellZTR08} & \makecell{$\mathcal{S}_{d q}=\left\{s_{q}: d \in s_{q}\right\}$} & \makecell{$\varepsilon_{r+1}=1$} \\
    \hline 
    DCM \cite{DBLP:conf/wsdm/GuoLW09} & \multirow{5}*{\makecell{$\mathcal{S}_{d q}=\left\{s_{q}: d \in s_{q}, r \leq l\right\}$ \\ $l$ is the rank of the\\ last-clicked document}} & \makecell{$\varepsilon_{r+1}=c_{r}^{(s)} \lambda_{r}+\left(1-c_{r}^{(s)}\right) \frac{\left(1-\alpha_{d q}\right) \varepsilon_{r}}{1-\alpha_{d q} \varepsilon_{r}}$ \\ with $\lambda_{r}=\frac{1}{\left|\mathcal{S}_{r}\right|} \sum_{s \in \mathcal{S}_{r}} \mathcal{I}(r \neq l)$} \\
    \cline{1-1}\cline{3-3}
    SDBN \cite{DBLP:conf/www/ChapelleZ09} & & \makecell{$\varepsilon_{r+1}=c_{r}^{(s)} \left(1-\sigma_{d q}\right)+\left(1-c_{r}^{(s)}\right) \frac{\left(1-\alpha_{d q}\right) \varepsilon_{r}}{1-\alpha_{d q} \varepsilon_{r}}$ \\ with $\sigma_{d q}=\frac{1}{\left|\mathcal{S}_{d q}^{\prime}\right|} \sum_{s \in \mathcal{S}_{d q}^{\prime}} \mathcal{I}\left(r^{(s)}=l\right)$ \\ where $\mathcal{S}_{d q}^{\prime}=\left\{s_{q}: d \in s_{q}, r_{d} \leq l, c_{d}^{(s)}=1\right\}$} \\
    \hline 
    \end{tabular}
\end{table*}

For a better illustration of how the continuation and satisfaction probabilities can be determined, Table \ref{tab:toy_example_clicks} provides a toy example with five sessions, for which we assume that the same ranking was logged for a single query $q$, where filled circles represent the clicks. For instance, we can determine the continuation probability of the second rank $\lambda_{r_2}$ by the sessions $s_1$, $s_3$, and $s_4$ at which the rank $r_2$ was clicked. For two out of these three sessions, the click at the second rank was followed by additional clicks at the lower ranks, which indicates that the users continued to browse through the ranking after having seen the document at rank $r_2$. Accordingly, the continuation probability is determined by this ratio, i.e., $\lambda_{r_2} = \frac{2}{3}$.

Similarly, we can determine the satisfaction probability at the second rank $\sigma_{d_{r_2} q}$. For one out of the three sessions ($s_4$), it was also the last click in the session. Accordingly, the satisfaction probability is determined by this ratio, i.e., $\sigma_{d_{r_2} q} = \frac{1}{3}$. Note that the continuation and satisfaction probabilities are complementary when comparing them for a single query, i.e., $\lambda_{r} = 1 - \sigma_{d q}$. The two click models DCM and SDBN differ if they are compared over multiple queries, as the continuation probability of DCM depends only on the rank $r$ and is determined over all queries. In contrast, the satisfaction probability of SDBN is specific to the query-document pair.

Suppose no clicks at a rank have been logged. In this case, it is impossible to determine the continuation and satisfaction probabilities (cf. $r_4$), and as a workaround, default probabilities can be used, or it is likewise possible to estimate values from the probability distribution.

\begin{table*}
    \caption{Toy example of the continuation $\lambda_{r}$ and satisfaction $\sigma_{d q}$ probabilities for five logged sessions for a single query $q$. The filled circles correspond to clicks.}
    \label{tab:toy_example_clicks}
    \begin{center}
     \begin{tabular}{| c || P{0.05\textwidth} | P{0.05\textwidth} | P{0.05\textwidth} | P{0.05\textwidth} | P{0.05\textwidth} || P{0.1\textwidth} | P{0.1\textwidth} |} 
     \hline
     \diagbox[width=3em]{$r_i$}{$s_i$} & $s_1$ & $s_2$ & $s_3$ & $s_4$ & $s_5$ & $\lambda_{r}$ & $\sigma_{d q}$ \\ 
     \hline\hline
     $r_1$ & \emptycirc & \emptycirc & \emptycirc & \fullcirc & \emptycirc & $\frac{1}{1} = 1.0$ & $\frac{0}{1} = 0.0$ \\ 
     $r_2$ & \fullcirc & \emptycirc & \fullcirc & \fullcirc & \emptycirc & $\frac{2}{3} = 0.\overline{6}$ & $\frac{1}{3} = 0.\overline{3}$ \\
     $r_3$ & \emptycirc & \fullcirc & \fullcirc & \emptycirc & \fullcirc & $\frac{1}{3} = 0.\overline{3}$ & $\frac{2}{3} = 0.\overline{6}$ \\
     $r_4$ & \emptycirc & \emptycirc & \emptycirc & \emptycirc & \emptycirc & - & - \\
     $r_5$ & \fullcirc & \emptycirc & \emptycirc & \emptycirc & \fullcirc & $\frac{0}{2} = 0.0$ & $\frac{2}{2} = 1.0$ \\
     \hline
     \end{tabular}
    \end{center}
\end{table*}

\subsection{Dataset}
\label{sec:dataset}

For our experiments, it is a fundamental requirement to have open data. Nowadays, several datasets are available for the general research community, but a large fraction of them is not suitable for our experiments. As pointed out before, previous work about click models was done in cooperation with large web search companies like Yahoo! \cite{DBLP:conf/sigir/CarteretteC18,DBLP:conf/kdd/LiAKMVW18,DBLP:conf/www/OzertemJD11,DBLP:conf/cikm/PiwowarskiZ07} or Yandex \cite{DBLP:conf/cikm/ChuklinSHSR13,DBLP:conf/clef/GrotovCMSXR15} and used entirely private or semi-public datasets. A popular dataset for the training of click models was made publicly available by Yandex as part of the \textit{Personalized Web Search Challenge}.\footnote{\url{https://www.kaggle.com/competitions/yandex-personalized-web-search-challenge/overview}} A similar dataset is publicly provided by Yahoo! as the \textit{L18 - Anonymized Yahoo! Search Logs with Relevance Judgments}.\footnote{\url{https://webscope.sandbox.yahoo.com/}} However, in both datasets, the web search results are anonymized, and no document collection of the entire corpus is provided. This is critical for our experiments as we want to build custom index and retrieval pipelines as defined above.

ORCAS \cite{DBLP:conf/cikm/CraswellCMYB20} is a companion dataset to MSMARCO that provides click-document pairs, and both the query as well as the document, are available in a clear text version. However, the DCM and SDBN click models do not only require triples containing the query, the documents, and the corresponding clicks but also the context of other documents in the SERP that were seen but not clicked, making ORCAS unusable for our experiments. We note that there exist several datasets that were curated in cooperation with the Chinese web search engine provider Sogou, like Sogou-QCL \cite{DBLP:conf/sigir/ZhengFLLZM18} or Sogou-SRR \cite{DBLP:conf/cikm/ZhangLMT18}, but these are not usable for us as non-Chinese speakers. More importantly, the dataset covers more general topics as it is based on web search results, but living labs usually have a domain-specific focus \cite{DBLP:journals/jdiq/JagermanBR18,DBLP:conf/clef/SchaerBCWST21a}.

Instead, we use the recently introduced TripClick \cite{DBLP:conf/sigir/RekabsazLSBE21} dataset of the biomedical search engine Trip in our experiments. It contains documents and user interaction logs covering a period of seven years, from 2013 to 2020. It was highlighted that the annotation coverage for the top results is low \cite{DBLP:conf/ecir/HofstatterASH22}, and recently, topical relevance judgments called TripJudge were introduced \cite{DBLP:conf/cikm/AlthammerHVH22}. In our experiments, we can only use data logs with information about the entire SERP, which are available from 13th August 2016. Furthermore, we restrict the sessions to the 50 most frequent queries in the dataset to ensure that at least 100 logged sessions are available for each query. We note that the Trip database has professional and non-professional users alike and that the head queries are a very particular query type. Even though they are domain-specific,  the selected sample of 50 queries is more generic than other queries in the torso or tail of the query distribution. Most of the query strings are composed of two terms only. As can be seen by the most frequent query in the logs (``\textit{covid-19 and pregnancy}''), the COVID-19 pandemic had an influence on the data logs, which is representational for the dynamic evaluation environment where new queries and information needs emerge from recent trends.

\subsection{Evaluation measures}
\label{sec:evaluation_measures}

In the following, we introduce the measures of our experimental evaluations, including the Log-Likelihood, the outcome of interleaving experiments, and the rank correlation measure Kendall's $\tau$. Besides, we include a preliminary system-oriented evaluation of the system rankings based on the TripClick and TripJudge relevance labels to verify the assumptions about the relative effectiveness of the single systems.

\subsubsection{Log-Likelihood}

This is a standard evaluation measure of click models, and it was found that better scores correlate with a higher fidelity of simulated clicks \cite{DBLP:conf/ictir/MalkevichMMR17}. We determine it over a run $R$ with $|\mathcal{Q}|$ queries and ranking length $n$ as follows:

\begin{equation}
    \mathcal{L L}(R)=\sum_{q \in \mathcal{Q}} \sum_{r=1}^{n} \log P\left(C_{d}=c_{d} \mid \mathbf{C}_{<r}\right)
\end{equation}

where $P\left(C_{d}=c_{d} \mid \mathbf{C}_{<r}\right)$ denotes the click probability of a particular click model for a document $d$ at rank $r$ given the ranking of a retrieval method for a query $q$ and the list of previous clicks $\mathbf{C}_{<r}$ before rank $r$ of the examined document. In our experiments, we use the TripClick data logs that contain SERPs with $20$ entries ($n=20$) and $|\mathcal{Q}| \in [1,50]$. Unlike previous work, we do not use Log-Likelihood to evaluate the click model itself but to distinguish between the ranking quality of retrieval systems. Assuming that a well-performing retrieval method delivers attractive rankings that result in clicks, the system maximizes the click probabilities, and thus the Log-Likelihood, over every result in a ranking list.

\subsubsection{Outcome of interleaving experiments}

Our interleavings are based on the Team Draft Interleaving algorithm \cite{DBLP:conf/cikm/RadlinskiKJ08}. The corresponding interleaved ranking lists can be decomposed into two sets containing the documents $D_{\mathrm{exp}}$ contributed by the experimental system and the documents $D_{\mathrm{base}}$ of the competing baseline. An experimental system wins if it contributes the document with the highest click probability to the interleaved ranking, i.e., we determine the rank of the document with the highest click probability by 
\begin{equation}
    r = \underset{k \in \left\{1, ..., n \right\}}{\arg\max} P\left(C_{k} \mid \mathbf{C}_{<k}\right)
    \label{eq:win}
\end{equation}
whereas we assign a \textit{win} if $d_r \in D_{\mathrm{exp}}$. Otherwise, the experimental system looses, i.e., $d_r \in D_{\mathrm{base}}$, and a \textit{loss} is assigned. Suppose the click probabilities of the interleaving are indifferent from those of a ranking with unknown documents. In that case, the click model cannot decide on a better system, and a \textit{tie} is assigned. Finally, the \textit{outcome} is determined over multiple queries $\mathcal{Q}$ and is defined as 
\begin{equation}
    \mathrm{Outcome} = \frac{\mathrm{Wins}_\mathcal{Q}}{\mathrm{Wins}_\mathcal{Q}+\mathrm{Losses}_\mathcal{Q}}
    \label{eq:outcome}
\end{equation}

A clear \textit{winner} achieves an outcome of $1.0$, whereas $0.5$ means that the experimental system is on par with the baseline, and any outcome below $0.5$ indicates an inferior experimental system.

\subsubsection{Relative system performance}
\label{sec:rel_sys}

As it is common practice when comparing relative system rankings, we use Kendall's $\tau$ to compare the reference system ranking $\mathcal{R}$ with the click model system ranking $\mathcal{R}^{\prime}$ as follows:

\begin{equation}
    \label{eq:ktau}
    \tau(\mathcal{R}, \mathcal{R}^{\prime}) = \frac{P-Q}{\sqrt{\big(P + Q + U\big)\big(P + Q + V\big) }}
\end{equation}

whereas $P$ is the total number of concordant pairs (system pairs that are ranked in the same order in both rankings), $Q$ is the total number of discordant pairs (system pairs that are ranked in the opposite order in the two rankings), $U$ and $V$ are the number of ties, in $\mathcal{R}$ and $\mathcal{R}^{\prime}$, respectively. As a rule of thumb, Voorhees considers correlations with $\tau > 0.9$ as acceptable \cite{DBLP:conf/sigir/Voorhees98}. We evaluate the system rankings resulting from the click model-based evaluations in reference to the LRM and IRM rankings, for which the relative orderings are motivated by the Tester-based approach (cf. \ref{sec:experimental_systems}). 

We note that there exist other measures to determine the correlation between the two rankings. For instance, the Rank-biased Overlap (RBO)~\cite{DBLP:journals/tois/WebberMZ10} can be used to quantify the overlap between two lists of ranked items. Opposed to Kendall's $\tau$, RBO does not require identical sets of ranked items to be compared, i.e., it can be used with rankings with infinite lengths and dissimilar sets of documents that may only overlap to some extent. Additionally, RBO models the user's browsing behavior by the transition probability $p$ to the next ranked item, which allows giving more weight to overlap in higher rank positions. The lower the transition probability $p$, the more emphasis is put on overlaps in higher-ranked positions, modeling an impatient user. While it is generally preferable to compare RBO alongside Kendall's $\tau$ when evaluating document rankings (where Kendall's $\tau$ is known as the stricter measure~\cite{DBLP:conf/sigir/Breuer0FMSSS20,DBLP:journals/ipm/MaistroBSF23}), we evaluate the relative system performance by Kendall's $\tau$ only for two reasons. First, we compare the ranking of systems, not documents, and there is no need to include a user model in the evaluation of relative system performance, as the user would not be exposed to the system but to their corresponding outputs --- the rankings. Second, we deal with a fixed set of systems, and Kendall's $\tau$ can be used for more rigorous evaluations.

\begin{figure}[!t]
    \includegraphics[width=\textwidth]{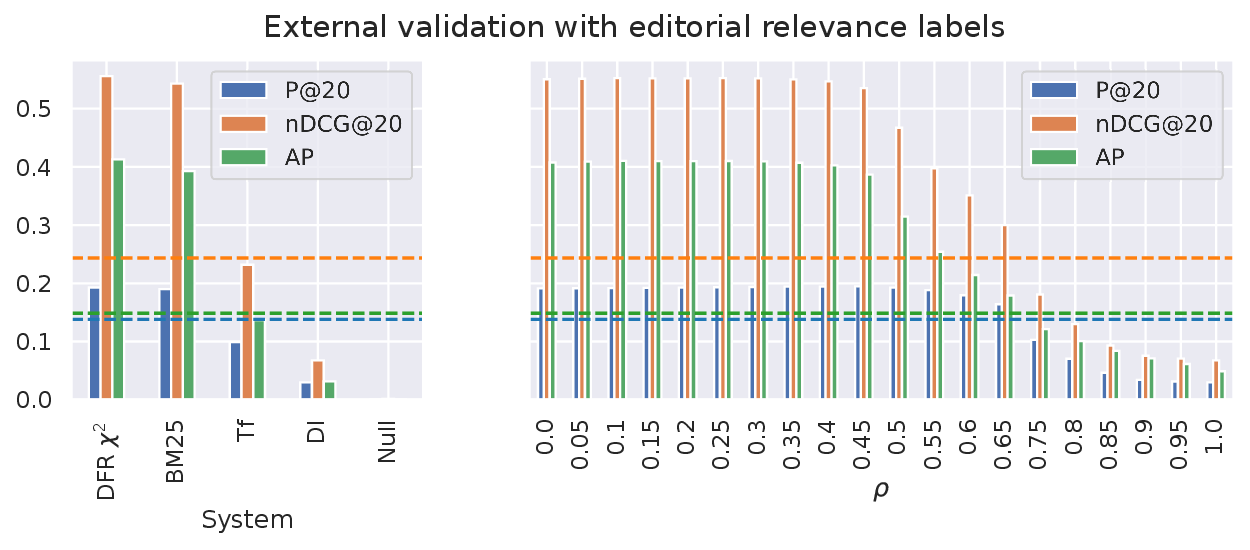}
    \caption{LRM (left) \& IRM (right) system rankings evaluated by editorial relevance judgments. The dashed lines correspond to the baseline system $\mathrm{IRM}_{\rho=0.7}$ (cf. \ref{sec:evaluation_measures}).}
    \label{fig:system_oriented_evaluations}
\end{figure}

In order to strengthen the reasoning behind the hypothesized system rankings, we evaluate them with the help of editorial relevance judgments. For this purpose, we use the previously mentioned TripJudge relevance labels \cite{DBLP:conf/cikm/AlthammerHVH22}. The results in Figure \ref{fig:system_oriented_evaluations} show that the system-oriented experiment gives evidence to the hypothesized relative orderings of the system performance. We can control the retrieval performance for both types of system constellations by choosing an entirely different ranking method or increasing the interpolation weight towards the inferior ranking criterion. 

Regarding the IRM ranking, we see that an interpolation parameter of $\rho \le 0.4$ does not substantially change the retrieval effectiveness. In our experiments, we set $\rho=\left\{ 0.4, 0.45, ... , 1.0 \right\}$, and we exclude all of the systems with $\rho \le 0.4$, as the experimental setup requires differences in effectiveness. Very likely, interpolations with $\rho \le 0.4$ do not impact effectiveness, as we determine the document's length by the abstracts. Naturally, abstracts are shorter than the corresponding full-texts, and abstracts do not differ in length as much as publications do. In the interpolations, the ranking method requires a certain weight to impact retrieval effectiveness. We consider the IRM system with $\rho=0.7$ as an adequate baseline, ranked in the middle of the IRM ordering with six systems performing better ($\rho < 0.7$) and six systems performing worse ($\rho > 0.7$). Similarly, $\mathrm{IRM}_{\rho=0.7}$ is almost on par with the Tf-based method that is in the middle of the LRM ranking.

In addition, we include Table \ref{tab:tripclick_vs_tripjudge} in the appendix, which compares the system-oriented measures P@20, nDCG@20, and AP based on the click-based relevance labels of TripClick to that based on the editorial relevance labels of TripJudge. As these results demonstrate, we can confirm that the coverage of relevant documents at the top ranks is higher when using the editorial TripJudge labels. However, in our case, both types of relevance labels agree about the relative system performance for both the LRM and IRM rankings. These system-oriented experiments are another perspective of the system performance, strengthening our methodology's reasoning as a form of external validation.

\subsection{Implementation details}
\label{sec:implementation}

We implement the experiments with the help of the Pyterrier retrieval toolkit \cite{DBLP:conf/ictir/MacdonaldT20} (the Python interface to the Java-based retrieval toolkit Terrier \cite{DBLP:conf/ecir/OunisAPHMJ05}) and the dataset library \texttt{ir\_datasets} \cite{DBLP:conf/sigir/MacAvaneyYFDCG21}, which features bindings to the TripClick dataset. We filter and select the session logs with the help of the NoSQL database MongoDB. We rely on the PyClick\footnote{\url{https://github.com/markovi/PyClick}} \cite{DBLP:series/synthesis/2015Chuklin} library when implementing the click models. In addition, we provide the required parsers to ingest the session logs from our database into the PyClick framework. All of the experiments are run on a Dell workstation with an \textit{Intel Xeon Gold 6144} CPU and 64 GB of RAM on \textit{Ubuntu 18.04 LTS}. The entire code to rerun the experiments is available on GitHub at \url{https://www.github.com/irgroup/validating-synthetic-usage-data}.

\section{Experimental Evaluations}
\label{sec:experimental_evaluations}

In the following, we present the experimental evaluations based on the analysis of the Log-Likelihood and the simulated interleaving experiments. In order to determine the performance of click models over an increasing amount of click data and queries, we randomly sample an increasing number of logged sessions, which are used to parameterize the click model. For each query $q \in \mathcal{Q}$, we randomly sample $s$ sessions ten times, i.e., we let the click model adapt to the given data sample (with $s$ sessions for $|\mathcal{Q}|$ queries) and evaluate the system rankings over ten trials. 

In the first experiment in \ref{sec:ll_eval}, the system rankings are determined by  Log-Likelihood based on the click probabilities, whereas in the second experiment in \ref{sec:outcome_eval}, the living labs are simulated, and the system rankings are based on the outcomes (cf. Equation \ref{eq:outcome}) of the corresponding interleaving experiments. Each system ranking that results from either the Log-Likelihood or the outcome measure is compared to the reference system rankings, which were introduced in \ref{sec:experimental_systems}, with the help of Kendall's $\tau$ (cf. Equation \ref{eq:ktau}).

\subsection{Log-Likelihood Evaluations}
\label{sec:ll_eval}

\begin{figure}[!t]
    \includegraphics[width=.8\textwidth]{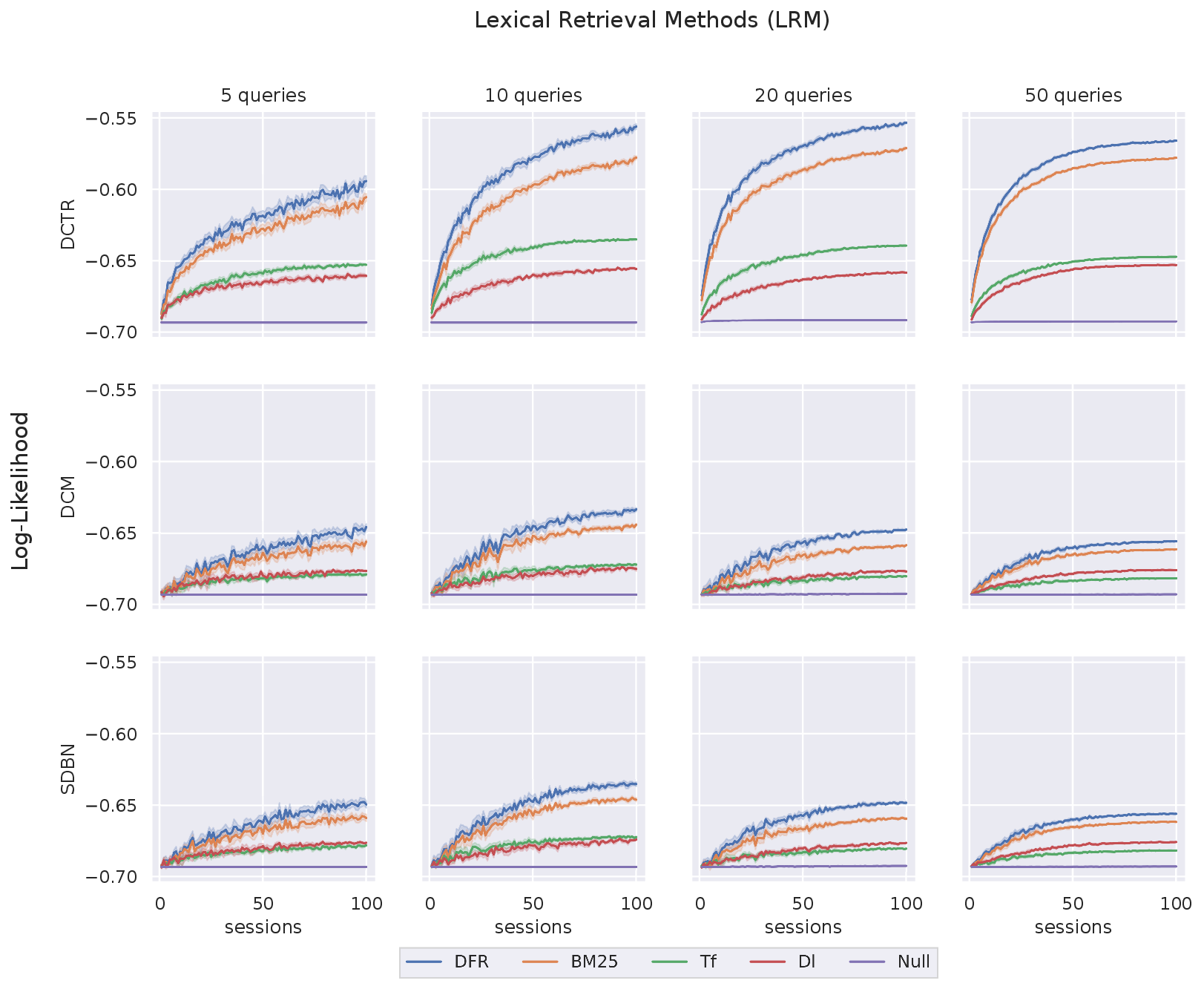}
    \includegraphics[width=.8\textwidth]{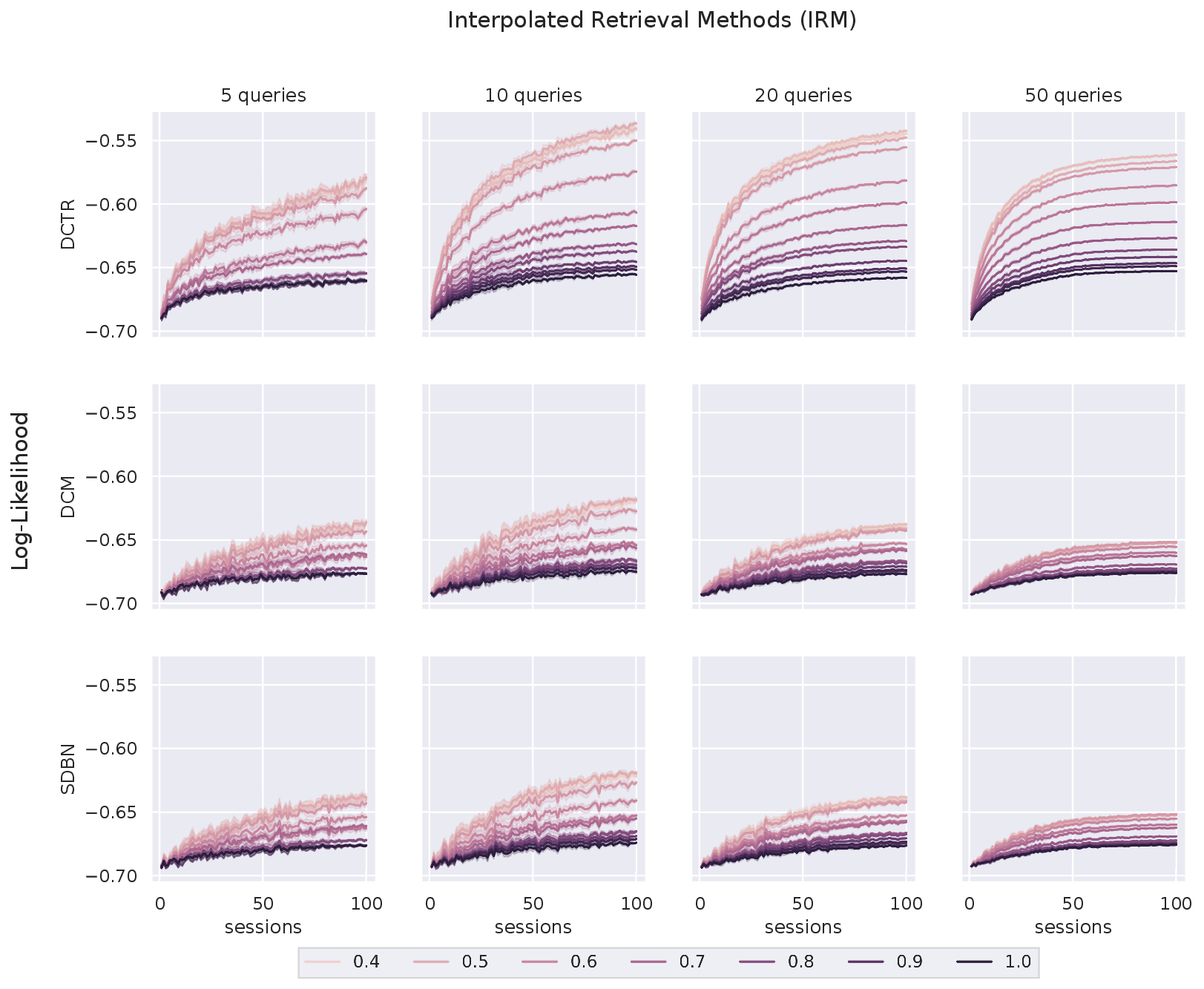}
    \caption{Log-Likelihood of the LRM and IRM system rankings based on the three click models DCTR, DCM, SDBN and compared by 5, 10, 20, and 50 queries.}
    \label{fig:ll}
\end{figure}

We determine the Log-Likelihood for all combinations resulting from the two system rankings and the three click models and evaluate them over an increasing amount of click log data that is used for parameterizing the click models. Figure \ref{fig:ll} shows the Log-Likelihood over the number of sessions with either 5, 10, 20, or 50 queries. Unsurprisingly, the Log-Likelihood increases as more sessions are used to parameterize the click models. As more click logs are available, the click models \textit{becomes familiar} with relevant, i.e., previously clicked documents, and consequently, there is a higher click probability.

There are apparent differences between DCTR and the other two click models when comparing them. In the case of the DCTR-based Log-Likelihood, the ranking order of documents is irrelevant as the click model does not account for the ranking position. Consequently, there is no rank-biased discount of the documents' attractiveness, leading to an overall higher Log-Likelihood of the DCTR model. In contrast, the document order affects the click probabilities of the DCM and SDBN click models, leading to an overall lower Log-Likelihood, which can be explained by the examination probabilities of these click models that are a rank-biased discount of the documents' attractiveness.

As it can be seen from the LRM ranking (in the upper half of Figure \ref{fig:ll}), the \textit{Null} system has a constant Log-Likelihood and is an estimate for lower bound performance. For the other systems, the Log-Likelihood increases as more sessions are considered, whereas the DFR and BM25 methods are quite distinct from the simple ranking criteria based on the term frequency (Tf) and document length (Dl). In the lower half of Figure \ref{fig:ll}, the IRM system rankings based on the Log-Likelihood aligns with the earlier system-oriented evaluations in \ref{sec:rel_sys}, i.e., the overall Log-Likelihood is lower (the retrieval system performs worse) when the interpolation parameter $\rho$ gives more weight to the inferior ranking criterion. 

By evaluating the Log-Likelihood with 50 queries, we see a steeper increase in the Log-Likelihood as more (possibly earlier clicked) documents are retrieved. Once enough click data are available, there are consistent click probabilities, as can be seen by the plateau-like shape of the Log-Likelihood plots with 50 queries. Any additional sessions with new click data only provide redundant relevance information and only affect the click probabilities to a negligible extent.  

In comparison, the Log-Likelihood averaged over fewer queries is noisier, as also can be seen by the larger confidence intervals, but it also increases over the sessions. By the example of the DCTR model, we see that the Log-Likelihood also increases as more queries are considered. However, comparing the results based on 10 or 20 queries to those based on 50 queries, there is a slightly higher Log-Likelihood when fewer queries are used. As the results are averaged over the queries, this can be explained by the higher click-through rates of the more frequent queries (top-10 or top-20), while less frequent queries also have lower click-through rates.  

\begin{figure}[!ht]
    \includegraphics[width=.925\textwidth]{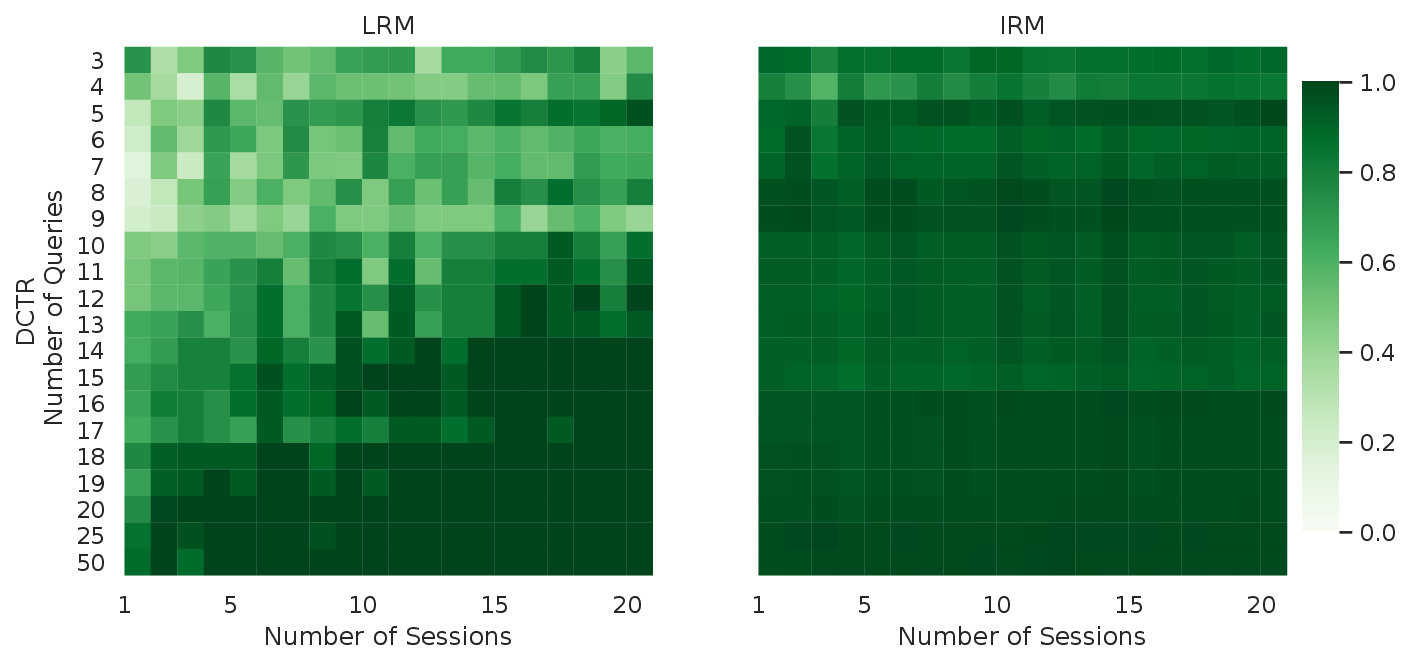}
    \includegraphics[width=.925\textwidth]{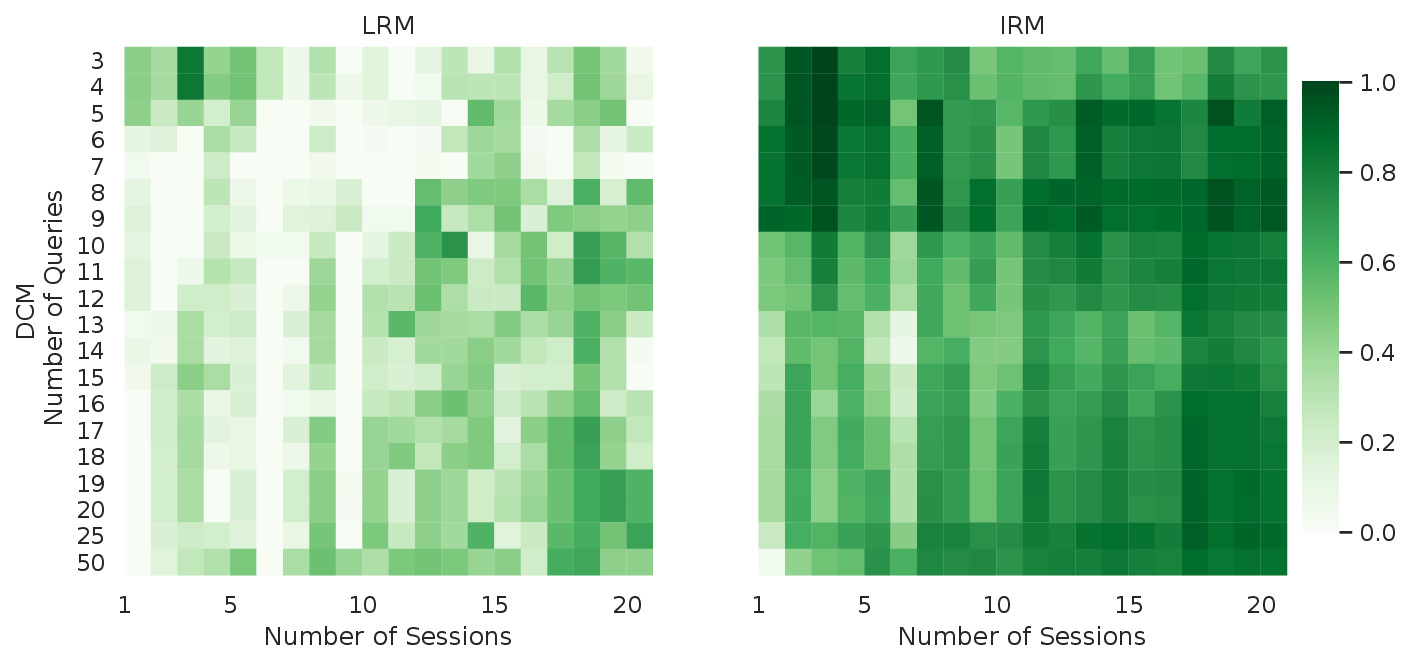}
    \includegraphics[width=.925\textwidth]{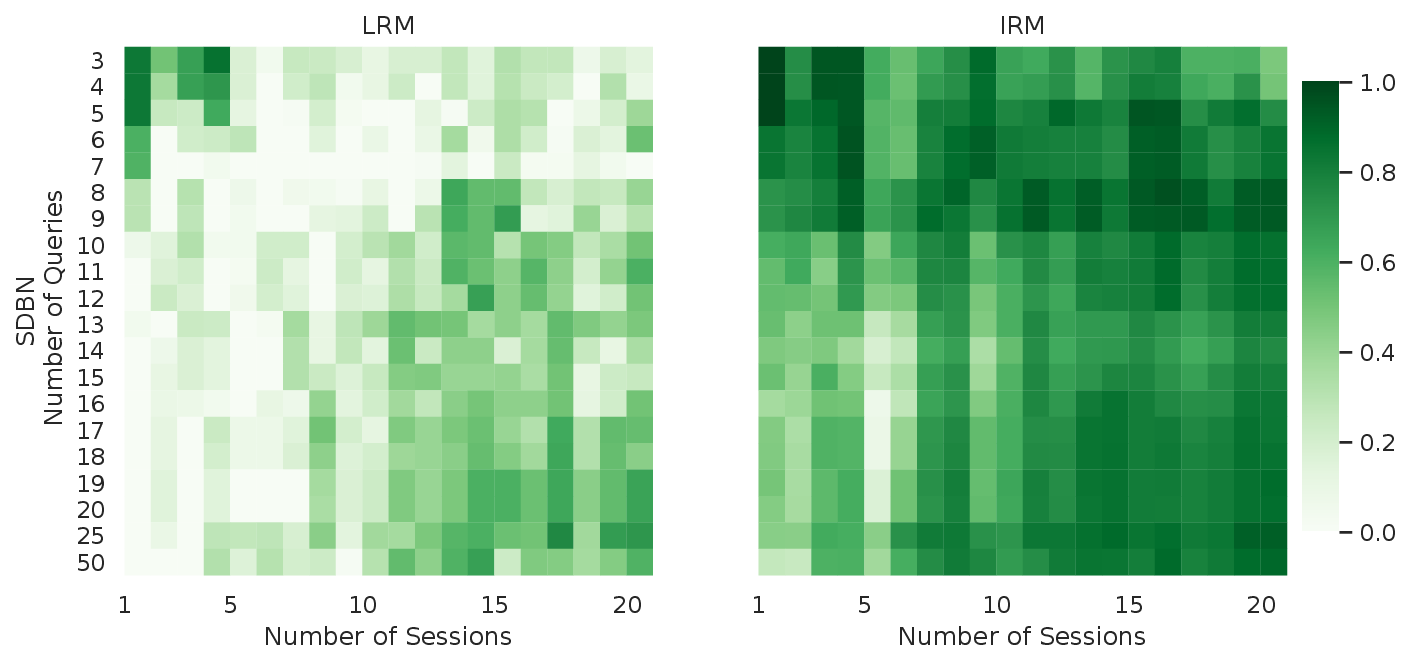}
    \caption{Kendall's $\tau$ of the LRM and IRM system rankings for different numbers of queries and logged sessions, compared for the three click models}
    \label{fig:ktau_ll}
\end{figure}

Overall, these preliminary evaluations suggest that either more queries or more sessions are required to distinguish between the single ranking systems. To this end, we conduct a more extensive analysis with an increasing number of queries and sessions. Figure \ref{fig:ktau_ll} compares Kendall's $\tau$ scores over different combinations of queries and log sessions for all three click models and the two system rankings. The heatmaps show the rank correlation in terms of Kendall's $\tau$ for the different combinations of queries (ranging from 3 to 50) and sessions (ranging from 1 to 20). The greener the corresponding patch, the higher the correlation between the reference and the click model system ranking.

The first heatmap based on DCTR and the LRM ranking shows a diagonal transition from the upper left corner to the lower right corner - the heatmap gets more greenish as more queries and sessions are used to evaluate the click model. In comparison, the IRM heatmap of the DCTR model has an overall darker appearance, which means that in comparison to the LRM ranking, less log data and queries are required in order to determine the correct system ranking. 

Evaluating 50 queries with a DCTR model based on 20 session logs for each query is already enough to reproduce the LRM system ranking with a perfect correlation of $\tau=1.0$. In contrast, the DCM and SDBN click models require more session logs to reliably reproduce the correct system orderings, resulting in lower correlation scores of $\tau_{\mathrm{DCM}}=0.4267$ and $\tau_{\mathrm{SDBN}}=0.5867$ on average with the same amount of queries and corresponding sessions. This can also be seen by the overall lighter heatmaps, which indicate low correlations between the system rankings.

In general, the IRM system ranking also results in higher Kendall's $\tau$ scores with fewer queries and sessions for DCM and SDBN, which suggests that it is easier for the click models to distinguish between systems that rely on the same retrieval method by the Log-Likelihood. We assume that the smaller document pool can explain this (cf. Figure \ref{fig:jacc}), i.e., there are fewer document candidates by which the method can be compared, and less click data are required for meaningful parameterization. 

We conclude that evaluating the relative system performance by the Log-Likelihood is a viable solution under the assumption that good-performing systems maximize the click-through rate only by the attractiveness of the ranking list. In comparison, DCTR is more robust and results in more reliable estimations when less log data are available. For instance, the LRM system rankings result in Kendall's $\tau$ scores of 1.0 with 50 queries and click data from 20 sessions for each query, while the Log-Likelihood based on DCM and SDBN scores is below $0.6$ when evaluated with the same amount of queries and click data. Overall, Log-Likelihood is lower when evaluated with the DCM and SDBN click models due to the examination probability discounting the attractiveness.

While the Log-Likelihood is an adequate indicator of system effectiveness in these evaluations, it is still an open question how it is related to user satisfaction or how it is related to editorial relevance. Additionally, the cognitive biases of the user should be considered in more user-oriented evaluations. Liu et al.~\cite{DBLP:conf/kdd/LiuMLZM19} found that user satisfaction is affected by the rank positions of relevant items. A large number of relevant items at the end of a session results in higher satisfaction than rankings of relevant items at higher positions earlier in the session. In this regard, DCTR attributes the same importance to documents irrespective of their position in the ranking. In comparison, the other click models rely on the cascade model that gives more weight to higher rank positions.

\subsection{Simulated Interleaving Experiments}
\label{sec:outcome_eval}

In the interleaving experiments, we determine the system ordering by the outcome measure (cf. Eq. \ref{eq:outcome}) for which the highest click probability is used as the winning criterion (cf. Eq. \ref{eq:win}). For each interleaving, the experimental ranking is interleaved with the baseline, which is consistent for both types of system rankings for the sake of better comparability and is set to $\mathrm{IRM}_{\alpha=0.7}$. 

\begin{figure}[!t]
    \includegraphics[width=\textwidth]{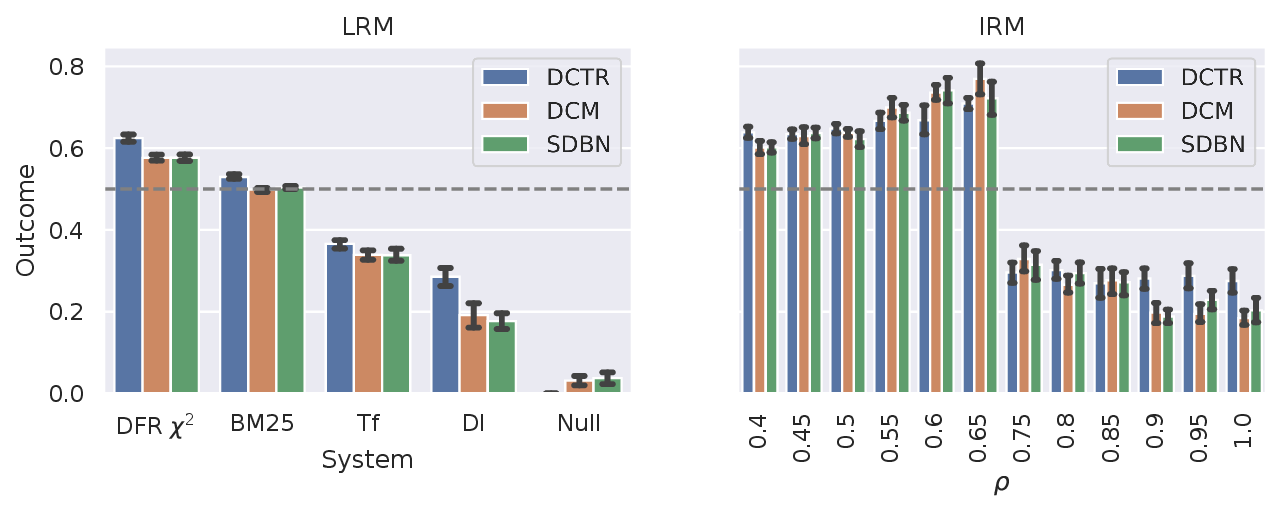}
    \caption{Outcome measures of interleaving experiments with click models based on 50 queries and 100 session logs. The dashed line corresponds to the baseline ($\mathrm{IRM}_{\rho=0.7}$) that is consistent for system rankings.}
    \label{fig:outcome_50queries_100sessions}
\end{figure}

Figure \ref{fig:outcome_50queries_100sessions} compares the outcomes for 50 queries with 100 session logs over ten trials for each experiment. Most strikingly, all of the click models can reproduce the correct orderings of the LRM system ranking, whereas, for the IRM system rankings, the relative ordering cannot be reproduced, but all of the click models can differentiate between systems that out- or underperform the baseline. In our analysis, often the \textit{winning} queries, i.e., those queries for which the experimental system wins, directly turn into losing queries as soon as the bad ranking criterion is assigned a higher weight than that of the baseline system.

\begin{figure}[!t]
    \centering
    \includegraphics[width=\textwidth]{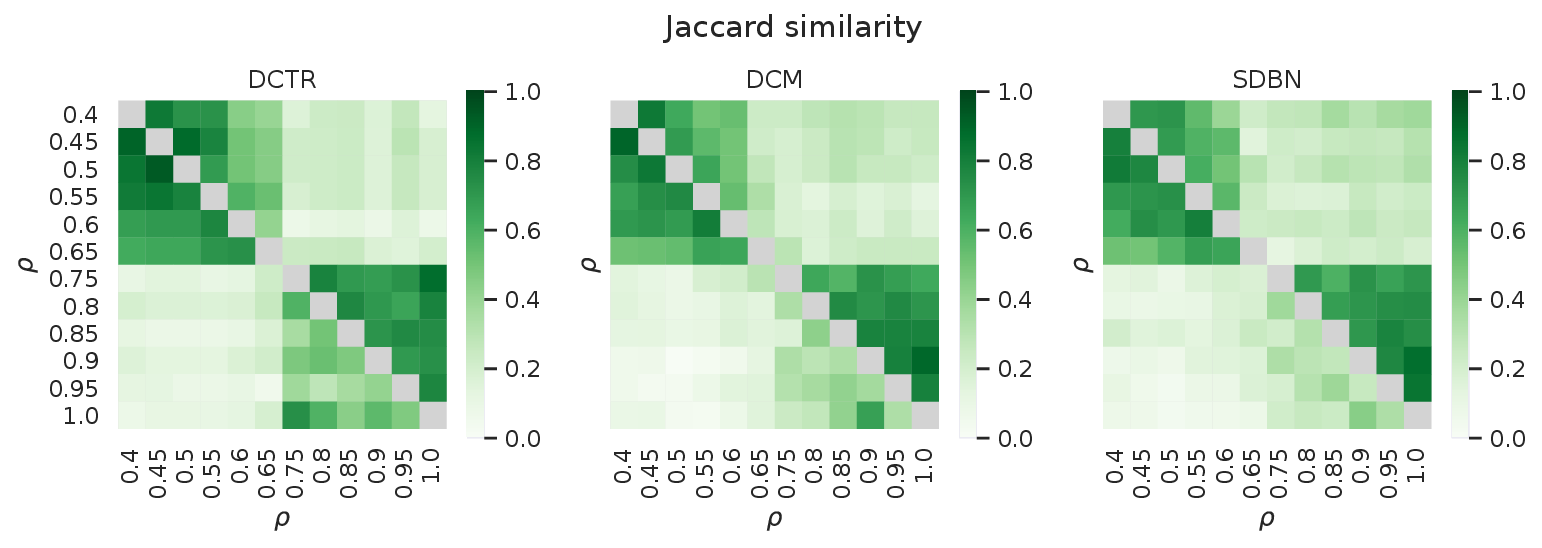}
    \caption{Jaccard similarity between the \textit{winning} (lower triangle) and \textit{losing queries} (upper triangle) of the simulated interleaving experiments with DCTR, DCM, SDBN click models.}
    \label{fig:jacc_queries}
\end{figure}

For better illustration, an in-depth analysis of the \textit{winning} and \textit{losing} queries is given in Figure \ref{fig:jacc_queries}. More specifically, the Jaccard similarity is shown for the \textit{winning} (lower triangle) and for the \textit{losing} (upper) queries over different interpolation weights, whereas winning and losing queries are those for which the experimental system is either assigned a \textit{win} or a \textit{loss}, respectively.

It can be seen that there are higher query similarities between those systems with an interpolation weight, which is either below or above that of the baseline system. However, there is a low overall similarity when comparing the winning/losing queries of system combinations with lower and higher interpolation weights (cf. to the light green areas in the lower left and upper right of the heatmap). This is independent of the click model, as the three heatmaps show similar results.

It means that for the IRM system ranking, the winning queries, i.e., those queries for which the experimental system wins, turn into losing queries as soon as the bad ranking criterion is assigned a higher weight than that of the baseline system. Queries resulting in \textit{ties} barely change, i.e., no or an equal number of clicks are made for both interleaved systems, as the click models cannot decide on a better system with unseen documents. These experimental results demonstrate that it can be problematic to compare systems with a small document pool with fewer document candidates and low click-through rates.

\begin{figure}[!t]
    \centering
    \includegraphics[width=\textwidth]{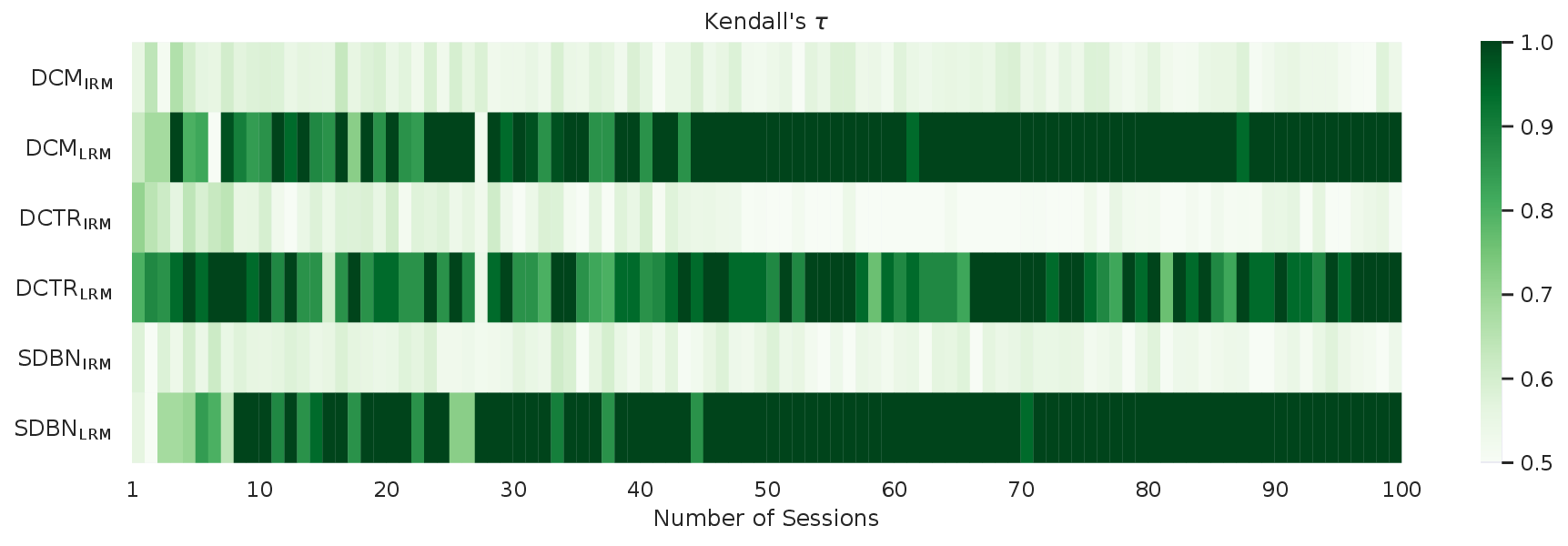}
    \caption{ Kendall’s $\tau$ of the LRM and IRM rankings based on simulated interleavings, compared for the click models DCTR, DCM, and SDBN parameterized with an increasing number of sessions.}
    \label{fig:outcome_vs_sessions}
\end{figure}

Finally, Figure \ref{fig:outcome_vs_sessions} shows Kendall's $\tau$ of the system rankings derived from the interleaving experiments resulting from click models parameterized over an increasing number of sessions. As can be seen by the light stripes in the heatmap, it is not possible to reproduce the correct ordering of IRM systems for any of the click models. Most of the rank correlations of the IRM rankings stay below $0.6$, which aligns with our earlier observations.

When comparing the LRM system rankings of the click models, we see that the DCTR model results in comparably higher correlations when less log data are available. For instance, the patches in the heatmap have a darker green when using 10 or fewer session logs per query for the DCTR model. However, the DCTR experiments demonstrate that the correlation scores do not stabilize even if more sessions are used for the parameterization. Once a certain amount of log data are used to parameterize the click models, DCM and SDBN deliver more robust correlation scores. For a better understanding and analysis, we determine the relative error between the cumulated and the ideal Kendall's $\tau$ score as

\begin{equation}
    \delta \tau = \frac{\Delta \tau}{\tau_{ideal}} = \frac{\tau_{ideal} - \tau_{sum}}{\tau_{ideal}} = 1 - \frac{\tau_{sum}}{\tau_{ideal}} = 1 - \frac{\sum_{s=1}^{\left|\mathcal{S}\right|} \tau_{s} }{\sum_{s=1}^{\left|\mathcal{S}\right|} 1 } = 1 - \frac{\sum_{s=1}^{\left|\mathcal{S}\right|} \tau_{s} }{\left|\mathcal{S}\right|}
    \label{eq:rel_err}
\end{equation}

where $\tau_{ideal}$ is considered as the sum of the ideal rank correlation up to the amount of considered sessions ${\left|\mathcal{S}\right|}$, and \textit{ideal} refers to a perfect rank correlation of 1. Accordingly, $\Delta \tau$ describes the difference between the actual sum of rank correlations and the ideal sum. A good performing user simulator or click model gives a low $\delta \tau$ score or minimizes it as it gets more session data for an adequate parameterization.

\begin{figure}[!t]
    \centering
    \includegraphics[width=\textwidth]{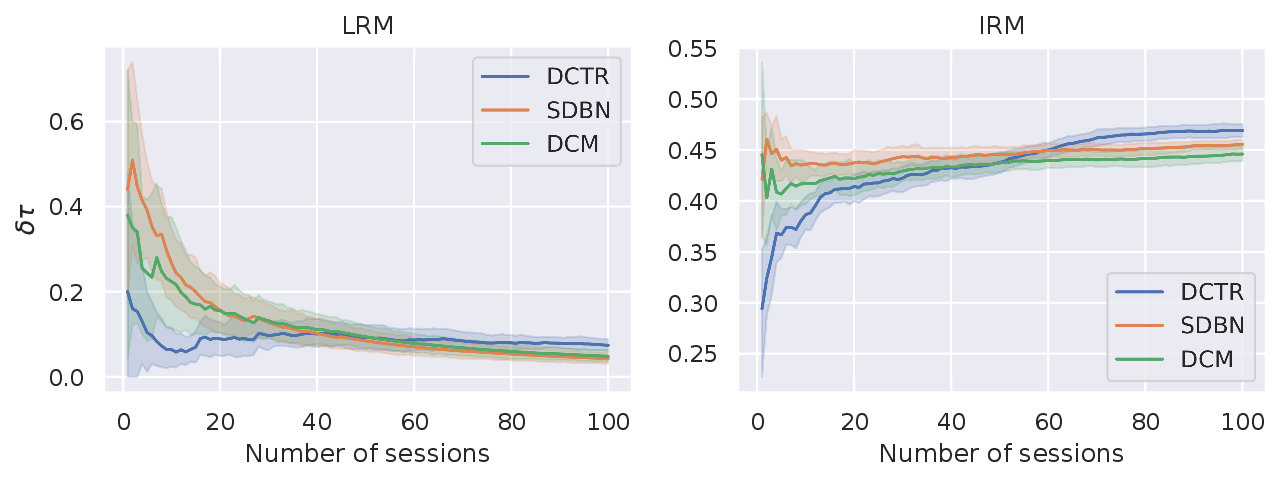}
    \caption{$\delta \tau$ over an increasing number of sessions for the LRM and IRM rankings based on interleavings, compared for the click models DCTR, DCM, and SDBN.}
    \label{fig:delta_tau}
\end{figure}

Figure \ref{fig:delta_tau} shows $\delta \tau$ for the click models in combination with both types of system rankings over an increasing amount of session logs. These results confirm that once enough session data are available, the DCM and SDBN click models can better distinguish between the relative system performance in these particular simulated interleaving experiments.

Regarding the LRM system ranking, there are higher errors for DCM and SDBN when only a few sessions are available, and the DCTR is a better choice when considering the lower error rates. However, it can be that with an increasing amount of click data, the error for both DCM and SDBN decreases while the error of the DCTR model evens out and does not decrease as more sessions are used for the parameterization.  

In comparison, it is generally harder for the click models to distinguish between the IRM system ranking based on interpolations. The experiments with 100 sessions result in considerably higher errors (higher $\delta \tau$ scores), but still, the DCM and SDBN give slightly better estimates than the DCTR. In this case, the $\delta \tau$ scores even out, while the scores of the DCTR still increase as more session logs become available. Similar to the earlier results, it is better to use DCTR when less log data are available. However, once enough logged clicks are available for the parameterization, the DCM and SDBN are less error-prone and more reliable.

\begin{figure}[!t]
    \centering
    \includegraphics[width=.52\textwidth]{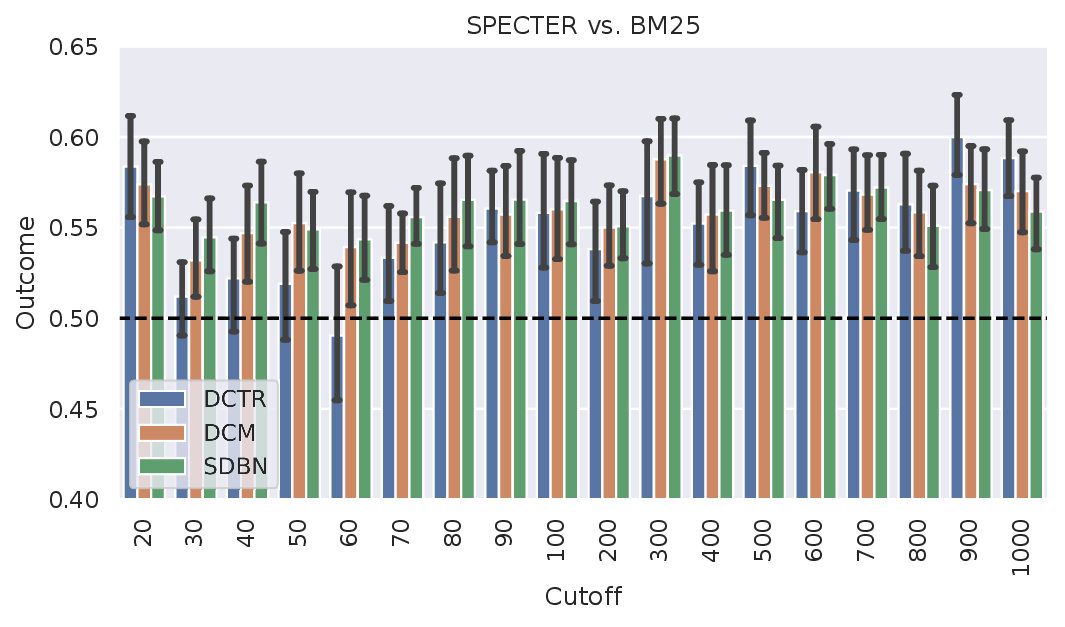}
    \includegraphics[width=.46\textwidth]{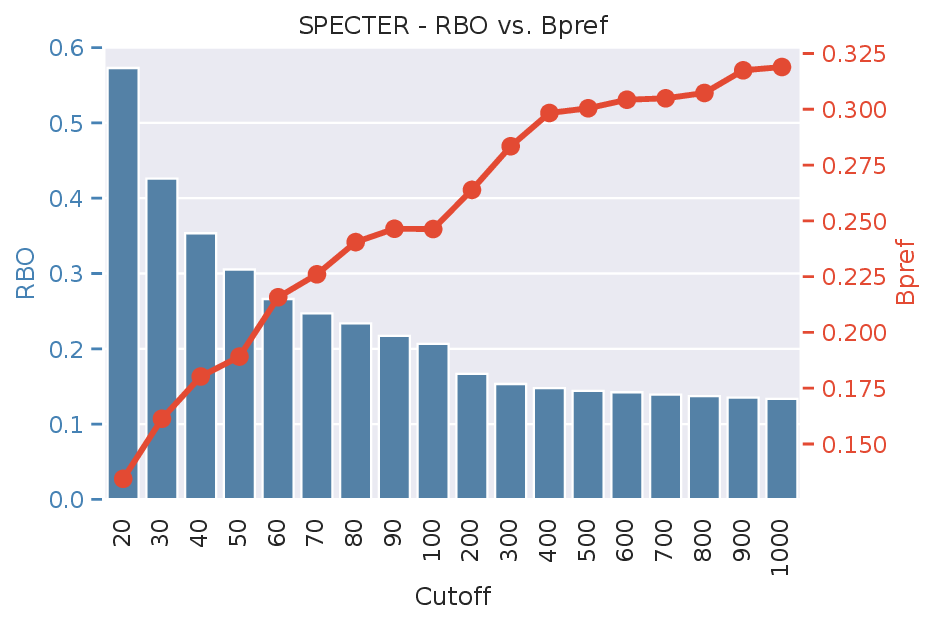}
    \caption{\textbf{Left:} Outcomes of simulated interleaving experiments with SPECTER-based rerankings of a first-stage BM25 ranking with different cutoff levels competing against the BM25 baseline. \textbf{Right:} The red point plot corresponds to the Bpref scores of the SPECTER rankings. The Rank-biased Overlap (RBO) with $p=0.95$ is determined between the first 20 documents of BM25 and SPECTER-based rerankings at different cutoffs.}
    \label{fig:specter}
\end{figure}

\subsection{Interleaving Experiments with Transformer-based Rankings}

In addition to the former experiments that confirmed the general plausibility of the introduced evaluation method, we demonstrate its application when evaluating state-of-the-art Transformer-based rankings. To have enough click logs available, our click models were parameterized with TripClick logs that are part of the data collection's training dataset. For this reason, it is not possible to fine-tune any Transformer-based method, as this would inevitably lead to leakage when using the same click logs during training and evaluation. As an alternative, we ground our experiments on the SPECTER language model~\cite{DBLP:conf/acl/CohanFBDW20,DBLP:journals/corr/abs-2211-13308} that we use as a zero-shot ranker without task-specific fine-tuning. More specifically, SPECTER generates dense vector representations of scientific documents. The language model is pretrained with the help of the documents' citation signals building upon SciBERT~\cite{DBLP:conf/emnlp/BeltagyLC19}, which, in turn, is a variant of the renowned BERT model~\cite{DBLP:conf/naacl/DevlinCLT19}. Cohan et al.~\cite{DBLP:conf/acl/CohanFBDW20} demonstrated that the model outperformed many baselines on different NLP tasks without fine-tuning. Likewise, the model performed well for zero-shot ad-hoc retrieval~\cite{DBLP:journals/corr/abs-2211-13308}.

We implement a typical two-stage ranking pipeline that includes a first-stage ranking based on BM25, reranked by SPECTER. Earlier experiments found that the reranking depth, i.e., the rank cutoff of the first-stage ranking, impacts the effectiveness of the final ranking~\cite{DBLP:series/synthesis/2021LinNY}. As the length of the first-stage ranking increases, the Transformer-based method can potentially find more relevant documents and push them to higher positions in the ranking. Conversely, more candidate documents result in higher computational costs, which can become critical in industrial applications, where system efficiency impacts user satisfaction. To this end, keeping the reranking depth low without sacrificing effectiveness is a desideratum.

In the simulated interleaving experiments, we let the final SPECTER rerankings with different cutoff levels compete against the BM25 baseline. In practice, this approach could be used to make estimates of an adequate reranking depth, considering effectiveness and efficiency tradeoffs. Figure~\ref{fig:specter} (left) shows the results of the simulated interleaving experiments. In addition, Figure~\ref{fig:specter} (right) shows Bpref~\cite{DBLP:conf/sigir/BuckleyV04}, which is a measure that solely considers judged documents, and the RBO~\cite{DBLP:journals/tois/WebberMZ10} between the first 20 documents of BM25 and SPECTER, which corresponds to the total number of documents shown to the click model. As we evaluate Bpref on the TripClick relevance judgments, it is a proxy measure of how well the system finds previously clicked documents. Similar to the evaluations of the previous subsection, we parameterize each click model with 100 sessions and simulate interleaving experiments with 50 head queries.

For the cutoffs at 30 to 60, the click models do not agree on the better-performing system. Based on the outcomes of the DCTR model, the reranking is indistinguishable from the BM25 baseline. As the cutoffs increase, the click models agree on SPECTER as the better-performing system; i.e., above a cutoff level of 70, SPECTER is considered more effective. In general, different click models make it feasible to evaluate the benefits for different kinds of user behaviors. If the click models disagree on the relative system performance, there is a higher risk of harming the users' search experience than in the case of agreement between the models.

With special regard to the cutoff level at 20, we see that the click models agree on SPECTER as the more effective system, which can be explained by the fact that, in this case, the SPECTER ranking is a purely (improved) reranking of the top 20 documents by BM25. For higher cutoffs, the reranker could bring up other documents --- unfamiliar to the click model --- among the first 20 ranking positions that are less likely to be clicked, which is the case for cutoff levels between 30 and 60. As the cutoff further increases above 60, SPECTER can rely on more relevant candidates in the BM25 ranking that are brought to the top 20 positions. 

This circumstance is further underlined by the fact that the Bpref scores increase, and the RBO scores decrease over the cutoffs. The increasing Bpref scores show that the SPECTER rerankings indeed benefit from an increase in the first-stage ranking's depth. Similarly, the RBO shows that increasing first-stage cutoffs leads to different document orderings in the top 20 positions. Even though the click models correctly identify SPECTER as the better system, a relative order is not evident from the outcome scores, i.e., there is no clear preference for any cutoff from 70 and above. 

By these results, we conclude that it is generally possible to identify better-performing Transformer rankings, e.g., in this particular case, having a reranking depth of at least 70 documents retrieved with BM25 is recommended. However, this experiment also demonstrates the limitations of the evaluation approach. It has to be considered that these Transformer-based rankings could bring up many documents that were not seen by the click models, which limits the estimates of the relative system effectiveness, especially for higher cutoff levels.

\section{Answers to the Research Questions}
\label{sec:answers}

In the following, we recapture our main findings of the experimental evaluations in the previous Section \ref{sec:experimental_evaluations} by giving answers to our three research questions.

\subsection{RQ1: Can click models reproduce system rankings?}

In our experimental evaluations, all click models can reproduce the system rankings if enough click logs are available, which is fundamental to our proposed methodology. We defined the simulation quality by how well the click model's click probabilities can reproduce the correct system ranking that is known in advance. The simulation quality improves depending on how much session data are available to parameterize the click model. In environments where user interaction data are sparse, keeping the required amount of user interaction data low becomes critical. In this regard, the DCTR model is able to distinguish reliably between the LRM systems by the Log-Likelihood with already 20 logged sessions if 50 queries are used in our experimental setup. In direct comparison, the IRM ranking can be reproduced with fewer data, which can be explained by a smaller pool of documents for which interaction data has to be logged.

\subsection{RQ2: Do continuation and satisfaction probabilities in click models improve the simulation quality?}

In our experimental setup, it is not recommended to use the DCM and SDBN for the Log-Likelihood in an interactive data-sparse setting. In the corresponding evaluations, DCM and SDBN result in overall lower scores in comparison to the DCTR model, which can be explained by the rank-biased discount of the attractiveness due to the examination probability. This is not critical when large amounts of session logs are available. For instance, if we can use 100 sessions per query, it is enough for adequate parameterization. However, compared to the DCTR, 20 sessions per query are not enough to let the DCM and SDBN reproduce the correct system ranking. On the other hand, the DCM and SDBN system rankings are a better choice when simulating the interleaving like they are implemented in living labs. In this case, the estimates of the LRM system ranking are much more robust, and the continuation and satisfaction probabilities of DCM and SDBN can indeed improve the simulation quality in our experimental setting.

\subsection{RQ3: How does the type of system ranking impact the outcomes of simulated interleaving experiments?}

While all of the models can determine the correct ordering of the LRM system ranking reasonably well in the simulated interleaving experiments, it is impossible to reproduce the correct IRM ranking. However, one can still distinguish between better and worse-performing IRM systems and separate them from the baseline. In our experimental setting, it is generally harder to reproduce the IRM ranking as there are deciding queries that either let the IRM system win or lose against the baseline system, depending on the interpolation weight. Once the interpolation parameter gives a higher weight to the bad ranking criterion, most of the queries, which formerly let the system win against the baseline, are the deciding queries that let the system lose against the baseline. This finding is critical for search platform operators, as different parameterizations of the same retrieval method may result in measurable differences in system-oriented experiments, while they are not reproducible in click model-based simulations.

\section{Discussion and Conclusion}
\label{sec:conclusions}

Living labs are a special type of human-in-the-loop environment that facilitates the evaluation of IR systems in real-world experiments. However, previous work has highlighted that user interaction data in living labs is usually sparse, and it is desirable not to damage a search platform's reputation with bad search results. These circumstances lead to the two requirements of 1) inferring relevance information from as little interaction data as possible and 2) keeping the online time of highly experimental systems short.

As a solution, it is possible to evaluate experimental systems with synthetic usage data based on simulations instead of risking the exposure of possibly poor results to real users. However, it remains unclear when a user simulator can be reliably used to simulate real user behavior by generating meaningful synthetic data. To this end, we introduced an evaluation approach for validating a click model's simulation quality in human-in-the-loop environments like living labs. 

Earlier living labs primarily logged user interaction data in the form of clicks that were used to evaluate the systems directly in interleaving experiments, but likewise, the clicks could be used to parameterize click models. However, it is often unclear if the click model received enough click logs for adequate parameterization. Our evaluation methodology aims at letting the click model decide about the relative system performance that is known with high confidence or based on some reasonable heuristics. In the literature, this approach was recently introduced as the Tester-based approach \cite{DBLP:conf/sigir/LabhishettyZ21,DBLP:conf/ecir/LabhishettyZ22}. The click model's system ranking is compared to the reference system ranking, and the rank correlation, determined by Kendall's $\tau$, is an indicator of the simulation quality.

In our experiments, we compared two different types of system rankings to validate the plausibility of the proposed evaluation method. The first ranking was composed of different lexical retrieval methods. In contrast, the second ranking was composed of a single ranking approach with different interpolations between a reasonable and less effective retrieval method. While these retrieval methods are rather simple compared to other state-of-the-art approaches, they are better candidates to validate the general plausibility of our approach. More specifically, the two types of system rankings cover the decision scenarios of platform operators. While the first system ranking corresponds to a scenario in which it is unclear what retrieval method to use in general, and a diverse set of methods should be evaluated, the second system ranking corresponds to a scenario in which a previously chosen retrieval method should be fine-tuned.

Our experiments have shown how the DCTR, DCM, and SDBN click models can be used in combination with the Log-Likelihood and the outcomes of simulated interleaving experiments for the assessments of retrieval methods and how much session data are required for reliable performance estimates. Overall, it is possible to reproduce the system rankings in simulations based on click models, confirming our methodology's general plausibility.

Regarding the evaluations based on the Log-Likelihood, the DCTR click model is a better choice if only a few sessions are logged. Our experiments showed that the DCTR could perfectly reproduce the system ranking with 20 logged sessions for 50 queries, while the DCM and SDBN could not. However, as more session logs become available, the DCM and SDBN click models are equally well-suited for this type of evaluation. While these outcomes are promising, it must be pointed out that the evaluation's focus is only on the attractiveness of the search results, which results in a simplified assumption about the users, making them more abstract. The rank-biased discount that better approximates real user behavior is not beneficial in this evaluation setting. 

This leaves the question of how the interpretation of the examination probabilities of the DCM and SDBN models is of benefit for the user simulations. For a better understanding, we simulated living lab experiments and let the click models decide about the preference for one of two competing systems in interleavings. The corresponding system rankings were based on the outcome measure and showed that, once again, DCTR is a better choice when only a small amount of session data are available. However, as more session logs became available, the DCM and SDBN gave better, i.e., more robust, estimates about the system rankings. 

When comparing the DCM to the SDBN model, there were no substantial differences in our experiments. The rank-biased discount of the DCM model is determined by a rank-dependent continuation probability, which is determined over all available sessions, while the SDBN introduces an additional satisfaction probability specific to the query-document pair. We conclude that for the underlying TripClick dataset, the consideration of the satisfaction probability did not make that much of a difference in comparison to the continuation probability.

We note that the decisions behind clicking on a snippet and annotating a document with a positive editorial label are fundamentally different. However, we think that click signal-based evaluations are a promising alternative when a curated test collection is not available, and click models can be used to evaluate the relative system performance when editorial relevance judgments are missing. For instance, click models could be used in a pre-assessment, similar to the idea of pseudo-relevance judgments \cite{DBLP:conf/sigir/SoboroffNC01}, to identify more promising systems for online experiments. Especially for small- and mid-scale search platforms that often partnered with living labs in the past, it would be a viable solution to use click signals instead of curating a costly test collection.

Finally, the simulated interleaving experiments with Transformer-based rankings revealed some limitations of the proposed methodology. More specifically, we compared rerankings based on SPECTER with different cutoff levels to the BM25 baseline ranking. While click models identified SPECTER as the more effective ranking method, it was impossible to derive a relative system ordering from the interleaving experiments. In principle, a higher cutoff level of the first-stage ranking should result in better retrieval performance, as the Transformer-based method can rely on more relevant items that are possibly reranked to higher positions in the list. However, our experiments showed that there is no preference for any cutoff level once the baseline ranking returns ranking lists with adequate depth. This circumstance can very likely be explained by the fact that SPECTER-based rerankings brought up many previously unclicked items, which are consequently unknown to the click model, still keeping an adequate amount of clicked items in higher positions to outperform the baseline. Generally, it is recommended to deploy different types of retrieval systems when collecting click feedback data, similar to relying on system diversity in the pooling when constructing a test collection. Nonetheless, the experiments could demonstrate how click models can at least be used to determine the required cutoff level. In practice, this method could help platform operators who aim for better estimates of the required cutoff level for balancing effectiveness and efficiency.  

Lastly, click data are biased \cite{DBLP:conf/sigir/White13}. To a certain extent, the click models address the bias that would emerge from using single clicks as relevance indicators, i.e., the probabilistic models grasp the behavior and preferences of the average user. However, there are other biases related to the click signals. For instance, a position or system bias was introduced by the unknown production system of the Trip database that we could not remove from the session logs. As part of future work, it should be analyzed to which extent these kinds of evaluations are insightful pre-assessments of the real system performance by deploying them in living labs \cite{DBLP:conf/cikm/GingstadJB20,DBLP:conf/clef/SchaerBCWST21a,DBLP:conf/clef/SchuthBK15}. In this way, the fidelity of the click models can be further investigated with real users.

\clearpage
\section*{APPENDIX}

\begin{table*}[!ht]
    \caption[]{System-oriented evaluations based on TripClick (click-based) and  TripJudge (editorial) relevance labels for the LRM and the IRM system rankings.}
    \label{tab:tripclick_vs_tripjudge}
    \centering
    \resizebox{\textwidth}{!}{
    \begin{tabular}{|p{1.2cm}||p{1.5cm}|p{1.5cm}|p{1.5cm}||p{1.5cm}|p{1.5cm}|p{1.5cm}|}
    \multicolumn{7}{c}{LRM} \\
    \multicolumn{1}{c}{}  & \multicolumn{3}{c}{TripClick} & \multicolumn{3}{c}{TripJudge} \\
    \hline 
    System & P@20 & nDCG@20 & AP & P@20 & nDCG@20 & AP \\
    \hline 
    \hline 
    DFR & 0.1439 & 0.1555 & 0.1404 & 0.1927 & 0.5557 & 0.4185 \\
    \hline 
    BM25 & 0.1337 & 0.1462 & 0.1363 & 0.1900 & 0.5429 & 0.3984 \\
    \hline 
    Tf & 0.0551 & 0.0542 & 0.0317 & 0.0989 & 0.2319 & 0.1420 \\
    \hline 
    Dl & 0.0470 & 0.0470 & 0.0265 & 0.0300 & 0.0675 & 0.0355 \\
    \hline 
    Null & 0.0010 & 0.0020 & 0.0005 & 0.0006 & 0.0015 & 0.0012 \\
    \hline 
    \multicolumn{7}{c}{}  \\
    \multicolumn{7}{c}{IRM} \\
    \multicolumn{1}{c}{}  & \multicolumn{3}{c}{TripClick} & \multicolumn{3}{c}{TripJudge} \\
    \hline 
    $\rho$ & P@20 & nDCG@20 & AP & P@20 & nDCG@20 & AP \\
    \hline
    \hline 
    0.00 & 0.1800 & 0.2363 & 0.1722 & 0.1912 & 0.5504 & 0.4075 \\
    \hline 
    0.05 & 0.1790 & 0.2354 & 0.1721 & 0.1915 & 0.5514 & 0.4089 \\
    \hline 
    0.1 & 0.1790 & 0.2352 & 0.1717 & 0.1919 & 0.5523 & 0.4102 \\
    \hline 
    0.15 & 0.1790 & 0.2351 & 0.1715 & 0.1921 & 0.5520 & 0.4100 \\
    \hline 
    0.2 & 0.1780 & 0.2341 & 0.1707 & 0.1926 & 0.5520 & 0.4098 \\
    \hline 
    0.25 & 0.1810 & 0.2370 & 0.1702 & 0.1932 & 0.5524 & 0.4100 \\
    \hline 
    0.3 & 0.1810 & 0.2365 & 0.1690 & 0.1936 & 0.5521 & 0.4093 \\
    \hline 
    0.35 & 0.1790 & 0.2324 & 0.1675 & 0.1942 & 0.5502 & 0.4070 \\
    \hline 
    0.4 & 0.1770 & 0.2307 & 0.1653 & 0.1944 & 0.5468 & 0.4026 \\
    \hline 
    0.45 & 0.1710 & 0.2279 & 0.1612 & 0.1946 & 0.5352 & 0.3867 \\
    \hline 
    0.5 & 0.1720 & 0.2045 & 0.1390 & 0.1926 & 0.4672 & 0.3144 \\
    \hline 
    0.55 & 0.1720 & 0.1965 & 0.1285 & 0.1884 & 0.3976 & 0.2542 \\
    \hline 
    0.6 & 0.1710 & 0.1862 & 0.1169 & 0.1795 & 0.3511 & 0.2142 \\
    \hline 
    0.65 & 0.1570 & 0.1677 & 0.1056 & 0.1640 & 0.2999 & 0.1791 \\
    \hline 
    0.7 & 0.1330 & 0.1435 & 0.0952 & 0.1380 & 0.2434 & 0.1486 \\
    \hline 
    0.75 & 0.1080 & 0.1208 & 0.0834 & 0.1027 & 0.1808 & 0.1214 \\
    \hline  
    0.8 & 0.0850 & 0.0952 & 0.0715 & 0.0704 & 0.1298 & 0.1006 \\
    \hline 
    0.85 & 0.0730 & 0.0840 & 0.0626 & 0.0463 & 0.0932 & 0.0840 \\
    \hline 
    0.9 & 0.0670 & 0.0784 & 0.0559 & 0.0342 & 0.0755 & 0.0712 \\
    \hline 
    0.95 & 0.0620 & 0.0737 & 0.0503 & 0.0314 & 0.0707 & 0.0613 \\
    \hline 
    1.0 & 0.0600 & 0.0717 & 0.0442 & 0.0300 & 0.0675 & 0.0489 \\
    \hline 
    \end{tabular}
    }
\end{table*}

\clearpage
\bibliographystyle{ACM-Reference-Format}
\bibliography{bibliography}

\end{document}